\documentclass[a4paper,11pt]{article}
\usepackage[left=2cm,right=2cm,top=3.5cm,bottom=3.5cm]{geometry}
\usepackage[dvipdfmx]{graphicx}
\usepackage{amsfonts}
\usepackage{physics,amsmath}
\usepackage{amssymb}
\usepackage{mathtools}
\usepackage{colortbl}
\usepackage{cite}
\usepackage{here}
\usepackage{hyperref}
\usepackage{mathrsfs}
\usepackage{algpseudocode}
\usepackage{xcolor}
\usepackage{ulem}
\numberwithin{equation}{section}
\usepackage{tabularx}
\usepackage{tabularray}
\usepackage{hhline}
\usepackage{makecell}
\usepackage{booktabs}
\usepackage{multirow}
\usepackage{subcaption}

\allowdisplaybreaks[1]
\hypersetup{
	colorlinks=true,
	linkcolor=ceruleanblue,
	filecolor=ceruleanblue,      
	urlcolor=ceruleanblue,
	citecolor=ceruleanblue,
}
\definecolor{ceruleanblue}{rgb}{0.0, 0.2, 0.6}
\renewcommand{\thefootnote}{*\arabic{footnote}}
\let\originalleft\left
\let\originalright\right
\renewcommand{\left}{\mathopen{}\mathclose\bgroup\originalleft}
\renewcommand{\right}{\aftergroup\egroup\originalright}

\newcommand{\de}{{\rm d}}

\newcommand{\blue}[1]{\textcolor{black}{#1}}

\newcommand{\comment}[1]{}

\newcommand{\fr}[2]{\frac{#1}{#2}}

\newcommand{\na}{\nabla}
\newcommand{\brb}[1]{\left[ #1 \right]}  
  
\newcommand{\be}{\begin{equation}}  
\newcommand{\ee}{\end{equation}}
\newcommand{\bem}{\begin{bmatrix}}
\newcommand{\eem}{\end{bmatrix}}
\newcommand{\Mpl}{M_{\rm Pl}}
\newcommand{\al}{\alpha}

\newcommand{\la}{\lambda}
\newcommand{\si}{\sigma}

\newcommand{\mn}{{\mu \nu}}

\newcommand{\mE}{\mathcal{E}}

\date{\today}

\begin{document}

\begin{flushright} {\footnotesize YITP-25-34, IPMU25-0010}  \end{flushright}

\begin{center}
\LARGE{\bf 
Inverting no-hair theorems: How requiring General Relativity solutions restricts scalar-tensor theories}
\\[1cm] 

\large{ Hajime Kobayashi$^{\,\rm a}$, Shinji Mukohyama$^{\,\rm a, \rm b}$, Johannes Noller$^{\,\rm c, \rm d}$, \\
Sergi Sirera$^{\,\bigstar, \rm a, \rm c, \rm d}$, Kazufumi Takahashi$^{\,\rm a}$, and Vicharit Yingcharoenrat$^{\,\rm b, \rm e}$}
\\[0.5cm]

\small{
 \textit{$^{\rm a}$
 Center for Gravitational Physics and Quantum Information, Yukawa Institute for Theoretical Physics, 
 Kyoto University, 606-8502, Kyoto, Japan}}
 \vspace{.2cm}

 \small{
 \textit{$^{\rm b}$
 Kavli Institute for the Physics and Mathematics of the Universe (WPI), The University of Tokyo Institutes for Advanced Study (UTIAS), The University of Tokyo, Kashiwa, Chiba 277-8583, Japan}}
 \vspace{.2cm}

 \small{
 \textit{$^{\rm c}$
 Department of Physics \& Astronomy, University College London, London, WC1E 6BT, U.K.}}
 \vspace{.2cm}

 \small{
 \textit{$^{\rm d}$
 Institute of Cosmology \& Gravitation, University of Portsmouth, Portsmouth, PO1 3FX, U.K.}}
 \vspace{.2cm}

 \small{
 \textit{$^{\rm e}$
 High Energy Physics Research Unit, Department of Physics, Faculty of Science, Chulalongkorn University, Pathumwan, Bangkok 10330, Thailand}}
 \vspace{.2cm}
\end{center}

\begingroup
\renewcommand{\thefootnote}{\relax} % Suppress numbering
\footnotetext{$\bigstar$ Corresponding author: sergi.sirera@port.ac.uk}
\endgroup

\vspace{0.3cm} 

\begin{abstract}\normalsize
Black hole solutions in general scalar-tensor theories are known to permit hair, i.e.~non-trivial scalar profiles and/or metric solutions different from the ones of General Relativity (GR). Imposing that some such solutions---e.g.~Schwarzschild or de Sitter solutions motivated in the context of black hole physics or cosmology---should exist, the space of scalar-tensor theories is strongly restricted.
Here we investigate precisely what these restrictions are within general quadratic/cubic higher-order scalar-tensor theories for stealth solutions, whose metric is given by that in GR, supporting time-dependent scalar hair with a constant kinetic term.
We derive, in a fully covariant approach, the conditions under which the Euler-Lagrange equations admit all (or a specific set of) exact GR solutions, as the first step toward our understanding of a wider class of theories that admit approximately stealth solutions. 
Focusing on static and spherically symmetric black hole spacetimes, we study the dynamics of linear odd-parity perturbations and discuss possible deviations from GR.
Importantly, we find that requiring the existence of all stealth solutions prevents any deviations from GR in the odd-parity sector.
In less restrictive scenarios, in particular for theories only requiring the existence of Schwarzschild(-de Sitter) black holes, we identify allowed deviations from GR, derive the stability conditions for the odd modes, and investigate the generic deviation of a non-trivial speed of gravitational waves.
All calculations performed in this paper are reproducible via companion {\texttt {Mathematica}} notebooks~\cite{ringdown-calculations}.
\end{abstract}

\vspace{0.3cm} 

\vspace{2cm}

\newpage
{
\hypersetup{linkcolor=black}
\tableofcontents
}

\flushbottom

\vspace{1cm}

%%%% -- Introduction -- %%%%%%%%%%%%%%%%%%%%%%%%%%%%%%%%%%%%%%%%%%%%%%%%%%%%%%%%

%%%%%%%%%%%%%%%%%%%%%%%%%%%%%%%%%%%%%%%%%%%%%%%%%%%%%%%%%%%%%%%%%%%%%%%%%%%%%
\clearpage
\section{Introduction}

The study of black hole solutions has been pivotal in advancing our understanding of gravity, offering a fertile testing ground for General Relativity (GR) and alternative theories of gravity. Among these alternatives, scalar-tensor (ST) theories have attracted significant attention due to their capacity to modify GR (potentially addressing some of the challenges in describing e.g.~dark energy, dark matter, or inflation) while maintaining a predictive and well-posed theoretical structure. In particular, the degenerate higher-order scalar-tensor (DHOST) theories~\cite{Langlois:2015cwa,Crisostomi:2016czh,BenAchour:2016fzp}, which generalise the Horndeski framework~\cite{Horndeski:1974wa,Deffayet:2011gz,Kobayashi:2011nu}, provide one of the richest known and well-defined landscapes for exploring deviations from GR.\footnote{There have been extensive studies on further generalisation of the extension of the DHOST framework.
One of them is to assume that the scalar field has a timelike gradient and to impose the degeneracy conditions only in the unitary gauge, which was dubbed U-DHOST~\cite{DeFelice:2018ewo,DeFelice:2021hps,DeFelice:2022xvq}.
Another is to perform a higher-derivative generalisation of invertible disformal transformations on Horndeski/U-DHOST theories~\cite{Takahashi:2021ttd,Takahashi:2022mew,Takahashi:2023jro,Takahashi:2023vva}, utilising the fact that an invertible transformation preserves the number of physical degrees of freedom~\cite{Domenech:2015tca,Takahashi:2017zgr}.}
The theories explored in this paper therefore represent some of the most comprehensive and general ST frameworks developed to date.\footnote{\blue{In this paper, we do not necessarily impose the degeneracy conditions; instead, we primarily work with \blue{generic higher-order ST} theories. The rationale for this choice is explained below.}}

In the context of such ST theories, a number of no-hair theorems exist, showing that (for large classes of ST theories) the presence of the extra scalar degree of freedom does not alter the background black hole solutions in the theory with respect to GR---see e.g.~\cite{Bekenstein:1995un,Mazur:2000pn,Hui:2012qt,Capuano:2023yyh}.
Black holes therefore retain their remarkable simplicity in these ST extensions of GR and (at the background level) are still fully specified by their mass and spin (and, if present, charge). However, in general ST theories so-called `hairy' black hole solutions exist, where the scalar has a non-trivial profile and, importantly, black hole solutions for the metric can also be affected by the presence of the scalar field. As such, general ST theories may also contain (some or all) no-hair black hole solutions as present in GR, as well as other solutions with hair. It is then interesting to turn the spirit of no-hair theorems on its head and ask the reverse question: If we require the presence of all, or some given specific, black hole solutions we are familiar with from GR, then 1)~to what extent does this restrict the theories in question, and 2)~how do the resulting restrictions affect the black hole dynamics, especially also the behaviour of perturbations around black holes as e.g.~observable via black hole ringdown? In this paper we will address these questions within the framework of cubic (and quadratic) \blue{higher-order scalar-tensor} (HOST) theories.
\blue{The motivation for focusing on theories that admit stealth solutions is threefold. First, from a phenomenological perspective, stealth solutions provide a compelling mechanism to evade current observational constraints on deviations from GR in strong-gravity regimes. By construction, these solutions allow for non-trivial scalar field configurations that leave the metric unchanged, making them compatible with existing observations while still allowing for novel scalar dynamics. Second, from a theoretical standpoint, the requirement that a theory admits stealth solutions acts as a non-trivial structural constraint on its operator content. This makes it a useful tool for carving out physically interesting subspaces within the larger theory space.
In particular, some of the ST theories studied for cosmology can also admit stealth configurations, potentially linking late-time scalar field dynamics to the strong-field regime in a coherent framework. Furthermore, as we shall explain below, one needs to go beyond DHOST by introducing so-called scordatura terms in order to avoid strong coupling of perturbations. This promotes a stealth solution to an
approximately stealth one that behaves as stealth for any practical
purposes at the level of the background and that is free from the
strong coupling issue for perturbations. Since scordatura terms are of
order unity (and not necessarily small) in the unit of the cutoff of
the theory, it is expected that the space of scalar-tensor theories
admitting approximately stealth solutions is significantly broader
than that admitting exact stealth ones. The present paper should be
considered as the first step towards the important problem of finding
such a broader theory space.
}

Black hole solutions in theories beyond GR that describe spacetimes indistinguishable from their GR counterparts (at the level of the background metric, but may be accompanied by a non-trivial scalar field profile, i.e.~they may still have `scalar hair') are typically referred to as `stealth' black hole solutions. 
For example, a shift-symmetric k-essence described by the action~$\int {\rm d}^4x\sqrt{-g}\,P(X)$, coupled to GR, admits stealth solutions if $P'(X)\equiv {\rm d}P(X)/{\rm d}X$ has a non-trivial root~$X=X_0\ne 0$, where $X\equiv -g^{\mn}\partial_{\mu}\phi\partial_{\nu}\phi/2$ and $g$ is the determinant of the metric. Indeed, for $X=X_0$, the stress-energy tensor~$T_{\mu\nu}=P'(X)\partial_{\mu}\phi\partial_{\nu}\phi+P(X)g_{\mu\nu}=P(X_0)g_{\mu\nu}$ is equivalent to that of an effective cosmological constant~$\Lambda_{\rm eff}=-P(X_0)/M_{\rm Pl}^2$ (with $\Mpl$ denoting the reduced Planck mass) which can be adjusted to any desired value by adding a bare cosmological constant to the k-essence action, meaning that any GR solution with or without the cosmological constant can be promoted to a stealth solution as far as the spacetime geometry admits a scalar profile with $X=X_0$. Actually, any regular spacetime locally admits a scalar profile with $X=X_0$: for $X_0>0$ (or $X_0<0$) one only needs to specify a spacelike (or timelike) hypersurface, to construct a congruence of geodesics orthogonal to the hypersurface and then identify $\phi=\sqrt{2|X_0|}\,\tau$ up to a constant shift, where $\tau$ is the proper time (or distance) along each geodesic.\footnote{\blue{
%This ensures $\phi_\mu$ is aligned with the geodesics, yielding $X=X_0$. While this is a local construction, it highlights that constant-$X$ profiles are not forbidden by geometry or kinematics. Whether such profiles correspond to global solutions of the equations of motion depends on the specific theory and the spacetime's global structure.
This ensures the relation between the scalar profile and the
proper time (or distance) coordinate of any Gaussian normal coordinate
system in any spacetime geometry. While the construction of the
Gaussian normal coordinate system is local as usual, a constant-$X$
profile is easily obtained everywhere in the spacetime region in which
a Gaussian normal coordinate system is defined. Once a Gaussian normal
coordinate system is globally specified in a spacetime region of
interest, the corresponding constant-$X$ profile of the scalar field
is obtained globally in the same spacetime region.}}
The stealth Schwarzschild solution of this type based on the Lema\^{i}tre coordinates was first found in \cite{Mukohyama:2005rw} and then later generalised to the stealth Schwarzschild-de Sitter (SdS) solution in Horndeski theory~\cite{Babichev:2013cya,Kobayashi:2014eva} and in DHOST theory~\cite{BenAchour:2018dap,Motohashi:2019sen,Takahashi:2019oxz,Minamitsuji:2019shy,Takahashi:2020hso}. It is even possible to obtain a stealth Kerr solution in DHOST theories~\cite{Charmousis:2019vnf,Takahashi:2020hso}. While perturbations around those stealth solutions suffer from strong coupling in the scalar sector~\cite{deRham:2019gha}, one can easily solve this problem and render those perturbations weakly coupled by taking into account higher derivative terms as in the effective field theory of ghost condensate \cite{Arkani-Hamed:2003pdi}, if and only if the scalar profile is timelike. This mechanism dubbed scordatura~\cite{Motohashi:2019ymr} was already taken into account in \cite{Mukohyama:2005rw} and the stealth Schwarzschild solution in k-essence was promoted to the approximately stealth solution in ghost condensate that behaves as stealth for any practical purposes in astrophysical scales at the level of the background~\cite{Cheng:2006us} and is free from the strong coupling issue for perturbations. For approximately stealth solutions in more general quadratic HOST with scordatura terms, see \cite{DeFelice:2022qaz}. In this way, once a stealth solution with a timelike scalar profile is found, one can easily promote it to an approximately stealth solution without the strong coupling issue. The timelike nature of such scalar profiles also offers the intriguing possibility that the associated time-dependence connects with cosmological dynamics driven by the scalar in the long distance limit, although it is of course non-trivial to have the same scalar yield leading order effects at long and short distance scales, see e.g. \cite{Noller:2019chl,Mukohyama:2024pqe}.

Cast in the above language, we are here investigating the effect of requiring the presence of stealth solutions in general ST theories.
More concretely, we consider solutions in ST theories that satisfy the Einstein equation in GR: $G_{\mu\nu} = \Mpl^{-2} T_{\mu\nu} - \Lambda g_{\mu\nu}$.
Here, $G_{\mu\nu}$ is the Einstein tensor, %$\Mpl$ denotes the reduced Planck mass, 
$T_{\mu\nu}$ is the stress-energy tensor of matter field(s), and $\Lambda$ is the (effective) cosmological constant.
Specifically, we will consider the effects of requiring the following stealth solutions:
\begin{enumerate}
    \item General stealth GR with 
    matter: {\it Any} metric satisfying $G_{\mu\nu} = \Mpl^{-2} T_{\mu\nu}-\Lambda g_{\mu\nu}$ is a solution.
    \item General stealth GR in vacuum: {\it Any} metric satisfying $G_{\mu\nu} =-\Lambda g_{\mu\nu}$ is a solution.
    \item Stealth SdS: The SdS metric~$g_{\mu\nu}^{\rm SdS}$ \eqref{eq:SdS} is a solution for any $\Lambda$.
    \item Stealth Schwarzschild: The Schwarzschild metric~$g_{\mu\nu}^{\rm Schw}$ \eqref{eq:Schw} is a solution.
\end{enumerate}
Solutions listed earlier include all of the later conditions listed---e.g.~requiring SdS solutions clearly includes requiring the presence of Schwarzschild solutions---so demanding the presence of the first set of solutions is a stronger requirement than for the second, which is stronger than for the third.

\textbf{Outline}: This work focuses on investigating the above stealth solutions in the context of HOST theories up to cubic order in double scalar derivatives (henceforth referred to as cubic HOST) with a linearly time dependent scalar field whose kinetic term is constant. We summarise the main findings in Table~\ref{main-table}.
The paper is organised as follows. 
In Section~\ref{sec:background} we introduce the cubic HOST action and derive the covariant equations of motion, showing how such equations of motion get simplified as we weaken the requirements on the form of the background metric.
%(i.e.~as we go from 1 to 4 in the above list).
We then derive the conditions that HOST functions need to satisfy in order to admit the existence of each class of exact stealth solutions.
In Section~\ref{sec:perturbations} we study the dynamics of odd-parity perturbations on a static and spherically symmetric background under the existence conditions.
We construct a second-order covariant Lagrangian collecting all possible contributions from HOST functions, and show which of them contribute in our setup. In order to diagnose potential departures from GR, we inspect the component form of the quadratic Lagrangian
%for odd-parity perturbations 
on S(dS) black holes.
In Section~\ref{sec:stab-conds} we determine stability conditions for perturbations and discuss the speeds of gravitational waves (GWs).
In Section~\ref{sec:genRWeq} we discuss an issue in deriving the master equation for odd-parity perturbations in non-shift-symmetric theories.
Finally, we summarise the paper and discuss several future directions in Section~\ref{sec:conclusions}.
All calculations performed in this paper are reproducible via 2 companion {\texttt {Mathematica}} notebooks~\cite{ringdown-calculations}, with calculations of Section \ref{sec:background} appearing in `Inverting-no-hair-theorems-I.nb', and those of Sections \ref{sec:perturbations} and \ref{sec:stab-conds} in `Inverting-no-hair-theorems-II.nb'. These notebooks construct an adaptable general formalism for the study of cubic HOST theories at the background and perturbative levels, which can be tuned to specific models.
\clearpage
\begin{table}[H]
  \centering
  \begin{tabular*}{\textwidth}{@{\extracolsep{\fill}} ccccc}
    \toprule
    \makecell{HOST \\ theory} & 
    \makecell{Stealth \\ metric} & 
    \makecell{Existence \\ conditions} & 
    \makecell{GR-deviations in \\ odd modes on S(dS)} & 
    \makecell{Stability \\ conditions} \\
    \midrule
    \multirow{4}{*}{Cubic} 
      & General GR with matter  & this work \eqref{eq:existcondGRT} & $\cross$ (this work) \eqref{eq:p-coefsGRmat} & $\checkmark$ \\
      & General GR vacuum & this work \eqref{eq:existcondGRV} & $\checkmark_1$ (this work) \eqref{eq:p-coefsGRvac} & \eqref{eq:stab-conds-GR-vac} \\
      & SdS           & this work \eqref{eq:existcondSdS} & $\checkmark_2$ (this work) \eqref{eq:p-coefsSdS} & \eqref{eq:stab-conds-group3} \\
      & Schwarzschild    & this work \eqref{eq:existcondSchw} & $\checkmark_3$ (this work) \eqref{eq:p-coefsSchw} & \eqref{eq:stab-conds-Schw} \\
    \midrule
    \multirow{4}{*}{\makecell{Shift-sym \\ cubic}} 
      & General GR with matter & this work \eqref{eq:existcondsGRmatSSCub} & $\cross$ \cite{Tomikawa:2021pca,Langlois:2022ulw} \eqref{eq:p-coefsGRmat} & $\checkmark$ \\
      & General GR vacuum & this work \eqref{eq:existcondsGRvacSSCub} & $\cross$ \cite{Tomikawa:2021pca,Langlois:2022ulw} \eqref{eq:p-coefsGroup1} & \eqref{eq:stab-conds-GR-vac-2} \\
      & SdS           & \cite{Minamitsuji:2019shy} \eqref{eq:existcondsSdSSSCub} & $\checkmark_1$ \cite{Tomikawa:2021pca,Langlois:2022ulw} \eqref{eq:coefsGroup2} & \eqref{eq:stab-conds-group4} \\
      & Schwarzschild    & \cite{Minamitsuji:2019shy} \eqref{eq:existcondsSchwSSCub} & $\checkmark_2$ \cite{Tomikawa:2021pca,Langlois:2022ulw} \eqref{eq:coefsGroup3} & \eqref{eq:stab-conds-group5} \\
    \midrule
    \multirow{4}{*}{Quadratic} 
      & General GR with matter  & \cite{Takahashi:2020hso} \eqref{eq:existcondsGRmatQuad} & $\cross$ (this work) \eqref{eq:p-coefsGRmat} & $\checkmark$ \\
      & General GR vacuum & \cite{Takahashi:2020hso} \eqref{eq:existcondsGRvacQuad} & $\checkmark_1$ (this work) \eqref{eq:p-coefsGroup1} & \eqref{eq:stab-conds-GR-vac-2} \\
      & SdS           & this work \eqref{eq:existcondsSdSQuad} & $\checkmark_2$ (this work) \eqref{eq:coefsGroup2} & \eqref{eq:stab-conds-group4} \\
      & Schwarzschild    & this work \eqref{eq:existcondsSchwQuad} & $\checkmark_2$ (this work) \eqref{eq:coefsGroup2} & \eqref{eq:stab-conds-group4} \\
    \midrule
    \multirow{4}{*}{\makecell{Shift-sym \\ quadratic}} 
      & General GR with matter  & \cite{Takahashi:2020hso} \eqref{eq:existcondsGRmatSSQuad} & $\cross$ \cite{Takahashi:2019oxz,Takahashi:2021bml} \eqref{eq:p-coefsGRmat} & $\checkmark$ \\
      & General GR vacuum & \cite{Takahashi:2020hso} \eqref{eq:existcondsGRvacSSQuad} & $\cross$ \cite{Takahashi:2019oxz,Takahashi:2021bml} \eqref{eq:p-coefsGroup1} & \eqref{eq:stab-conds-GR-vac-2} \\
      & SdS           & \cite{Motohashi:2019sen} \eqref{eq:existcondsSdSSSQuad} & $\checkmark_1$ \cite{Takahashi:2019oxz,Takahashi:2021bml} \eqref{eq:coefsGroup2} & \eqref{eq:stab-conds-group4} \\
      & Schwarzschild   & \cite{Motohashi:2019sen} \eqref{eq:existcondsSchwSSQuad} & $\checkmark_1$ \cite{Takahashi:2019oxz,Takahashi:2021bml} \eqref{eq:coefsGroup2} & \eqref{eq:stab-conds-group4} \\
    \bottomrule
  \end{tabular*}
  \caption{This table organises the main results of this paper as well as previous literature. We classify different combinations of theories and stealth metric solutions. The third column collects the works where the conditions ensuring the existence of the corresponding stealth solutions were first derived. The fourth column indicates whether GR deviations are present in each case in linear odd-parity perturbations on stealth S(dS) solution, and the subscript in $\checkmark$ counts the number of independent combinations of beyond-GR parameters present.
  (See Table~\ref{beyond-GR-params-table} for a summary of the beyond-GR parameters for each case.)
  The final column refers to the stability conditions of each case, with a $\checkmark$ denoting that stability conditions are automatically satisfied under the existence conditions.
  In the main body of the paper we use a simplified nomenclature for the different cases,
  namely $\prescript{\text{symmetry}}{}{\textbf{Theory}}_{\text{solution required}}$, so e.g., $\prescript{\text{SS}}{}{\textbf{Cubic}}_{\text{SdS}}$ for shift-symmetric cubic HOST theories where we impose the conditions requiring the existence of stealth SdS solutions.
  }
  \label{main-table}
\end{table}
\clearpage
\section{Field equations and existence conditions for stealth GR solutions}\label{sec:background}
\subsection{Quadratic/cubic HOST theories}\label{sec:HOST}
We examine the following action composed of the metric~$g_{\mu\nu}$ and the scalar field~$\phi$~\cite{Langlois:2015cwa,Crisostomi:2016czh,BenAchour:2016fzp}:
\begin{align}
    S_{\rm grav}=\int {\rm d}^4x\sqrt{-g}\biggl[
    &F_0(\phi,X)+F_1(\phi,X)\Box\phi+F_2(\phi,X) R
    +\sum_{I=1}^5A_I(\phi,X)L^{(2)}_{I} \nonumber \\
    &\quad\quad\quad\quad\quad+F_3(\phi,X)G_{\mu\nu}\phi^{\mu\nu}
    +\sum_{J=1}^{10}B_J(\phi,X)L_J^{(3)}
    \biggr] + \int {\rm d}^4x\sqrt{-g}\,L_{\rm m}\;, \label{eq:action1}
\end{align}
where $L_{\rm m}$ is the matter Lagrangian (assumed to be minimally coupled to gravity)\footnote{%\blue{Theories with minimally coupled matter are sometimes referred to as `metric theories', see e.g. \cite{Zosso:2024xgy}.}
\blue{Theories where freely falling bodies follow geodesics of a (symmetric) metric are sometimes referred to as `metric theories'~\cite{Will_2018}, which is the case in our setup.}}, $X \equiv -\phi_\mu\phi^\mu/2$,
$\phi_\mu \equiv \nabla_\mu\phi$, and $\phi_{\mu\nu} \equiv \nabla_\nu\nabla_\mu\phi$.
$R$ and $G_{\mu\nu}$ are the %4d
four-dimensional Ricci scalar and the Einstein tensor, respectively.
$L_I^{(2)}$ and $L_J^{(3)}$ comprise all possible terms built from $\phi_\mu$ and $\phi_{\mu\nu}$ which are quadratic and cubic in $\phi_{\mu\nu}$, respectively, and are written explicitly as
\begin{equation}
   \begin{aligned}
    &L_1^{(2)}=\phi_{\mu\nu}\phi^{\mu\nu} \;, & & L_{1}^{(3)} = (\Box \phi)^{3} \;, &  
    &L_{2}^{(3)} = (\Box \phi)\phi_{\mu \nu}\phi^{\mu \nu}\;,&  \\
    &L_2^{(2)}=(\Box\phi)^2,& &L_{3}^{(3)} = \phi_{\mu \nu} \phi^{\nu \rho} \phi^{\mu}_{\rho}\;,& 
    &L_{4}^{(3)} = (\Box \phi)^{2} \phi_{\mu} \phi^{\mu \nu} \phi_{\nu}\;,& \\
    &L_3^{(2)}=(\Box \phi)\phi^{\mu} \phi_{\mu \nu}\phi^{\nu}\;,& &L_{5}^{(3)} = \Box \phi \phi_{\mu} \phi^{\mu \nu} \phi_{\nu \rho}\phi^{\rho}\;,&
    &L_{6}^{(3)} = \phi_{\mu \nu} \phi^{\mu \nu} \phi_{\rho} \phi^{\rho \sigma} \phi_{\sigma}\;,&  \\
    &L_4^{(2)}=\phi^{\mu}\phi_{\mu \rho}\phi^{\rho \nu}\phi_{\nu}\;,& &L_{7}^{(3)} = \phi_{\mu} \phi^{\mu \nu} \phi_{\nu \rho} \phi^{\rho \sigma} \phi_{\sigma}\;,&
    &L_{8}^{(3)} = \phi_{\mu} \phi^{\mu \nu} \phi_{\nu \rho} \phi^{\rho} \phi_{\sigma} \phi^{\sigma \lambda} \phi_{\lambda}\;, & \quad  \\
    &L_{5}^{(2)}=(\phi^{\mu}\phi_{\mu \nu}\phi^{\nu})^{2},& &L_{9}^{(3)} = \Box \phi (\phi_{\mu} \phi^{\mu \nu} \phi_{\nu})^{2}\;,&
    &L_{10}^{(3)} = (\phi_{\mu} \phi^{\mu \nu} \phi_{\nu})^{3}\;.
\end{aligned} 
\end{equation}

The action introduced above encompasses both standard Horndeski~\cite{Horndeski:1974wa,Deffayet:2011gz,Kobayashi:2011nu} and beyond-Horndeski/DHOST~\cite{Zumalacarregui:2013pma,Gleyzes:2014dya,Gleyzes:2014qga,Langlois:2015cwa,Crisostomi:2016czh,BenAchour:2016fzp} theories as particular limits. For instance, the Horndeski action in the standard form with the Galileon functions can be recovered with the choices:
\begin{align}
\begin{split}
    &F_0=G_2\;, \quad F_1=G_3\;, \quad F_2=G_4\;,\quad F_3=G_5\;,\quad A_1=-A_2=-G_{4X}\;,\\
    &6B_1=-2B_2=3B_3=-G_{5X}\;, %\quad A_I=B_J=0 \;,
\end{split}
\end{align}
and $A_I=B_J=0$ for $I=3,4,5$ and $J=4, \cdots,10$.
The full classification of quadratic/cubic DHOST theories can be found in \cite{BenAchour:2016fzp}, and there are a large number of subclasses distinguished by different sets of degeneracy conditions.
Among these subclasses, there is one that can be obtained from the Horndeski class via invertible conformal/disformal transformation, which was called ``disformal Horndeski'' class in \cite{Takahashi:2022mew}.\footnote{It was shown in \cite{Langlois:2017mxy} that in the context of cosmology the disformal Horndeski class is phenomenologically favoured, unlike other subclasses of DHOST theories.}
(Note that an invertible conformal/disformal transformation preserves the number of physical degrees of freedom~\cite{Domenech:2015tca,Takahashi:2017zgr}.)
In the terminology of \cite{BenAchour:2016fzp}, this class corresponds to a sum of the quadratic DHOST of class~${}^{2}$N-I and the cubic DHOST of class~${}^{3}$N-I.
For the quadratic part (characterised by $F_2$ and $A_I$'s), the degeneracy conditions are given by\footnote{Note that degeneracy conditions can be imposed at the level of Lagrangian (i.e.~before specifying a background).}
    \begin{align}
    \begin{split}
	A_2&=-A_1\;, \\
        A_4&=\fr{(3F_2+8XA_1+2X^2A_3)(A_1+XA_3+F_{2X})^2-2A_3(3XA_1+X^2A_3+F_2+XF_{2X})^2}{2(F_2+2XA_1)^2}\;, \\
	A_5&=\fr{(A_1+XA_3+F_{2X})\brb{A_1(A_1-3XA_3+F_{2X})-2A_3F_2}}                 {2(F_2+2XA_1)^2}\;,
    \end{split}\label{DC_quadratic}
    \end{align}
where $F_2\,(\ne 0)$, $A_1$, and $A_3$ are free functions and the condition~$F_2+2XA_1\ne 0$ is assumed.
For the cubic HOST characterised by $F_3$ and $B_J$'s, we have
    \begin{align}
    &-\fr{B_2}{3}=\fr{B_3}{2}=B_1\;, \quad
    B_5=-B_7=\fr{4XB_4F_{3X}-(6B_1+F_{3X})^2}{12XB_1}\;, \quad
    B_8=\fr{B_5(4XB_4-6B_1-F_{3X})}{12XB_1}\;, \nonumber \\
    &B_6=-B_4\;, \quad
    B_9=\fr{B_4(4XB_4-6B_1-F_{3X})}{6XB_1}\;, \quad
    B_{10}=\fr{B_4(4XB_4-6B_1-F_{3X})^2}{24X^2B_1^2}\;, \label{DC_cubic}
    \end{align}
where $F_3$, $B_1\,(\ne 0)$, and $B_4$ are arbitrary functions.
Moreover, when both the quadratic and cubic parts are present, one has to impose the following conditions on top of Eqs.~\eqref{DC_quadratic} and \eqref{DC_cubic}:
    \begin{align}
    \begin{split}
    A_3&=\fr{A_1(4XF_{2X}-3F_2)}{XF_2}+\fr{XF_{2X}-F_2}{X^2}-\fr{F_{3X}(F_2+2XA_1)^2}{6X^2F_2B_1}\;, \\
    B_4&=\fr{6B_1(F_2-XF_{2X})+F_{3X}(F_2+XA_1)}{2XF_2}\;.
    \end{split}\label{DC_quadratic+cubic}
    \end{align}

Having said that, in what follows, we do not necessarily impose the degeneracy conditions for the following reason.
When we consider perturbations about stealth solutions (i.e., those with the metric of GR solutions) with a timelike scalar profile, the perturbations would be strongly coupled in the asymptotic Minkowski/de Sitter region~\cite{Motohashi:2019ymr}.
Therefore, in order to render those perturbations weakly coupled, one has to take into account deviation from the degeneracy conditions, which is known as the scordatura mechanism~\cite{Motohashi:2019ymr}.
Of course, any deviation from the degeneracy conditions leads to the appearance of an Ostrogradsky ghost in general, and therefore we assume the deviation is tiny so that the mass of the Ostrogradsky ghost is heavy enough.
(An exception is U-DHOST, where the scordatura mechanism is implemented by default while the Ostrogradsky ghost is intrinsically absent~\cite{DeFelice:2022xvq}.)
Note that we will mainly focus on odd-parity perturbations about static and spherically symmetric background, where the problems of the strong coupling and Ostrogradsky ghost are irrelevant.\footnote{Note also that there is a special class of ST theories where the scalar field does not propagate, e.g.~(extended) cuscutons~\cite{Afshordi:2006ad,Iyonaga:2018vnu}, and the odd-parity perturbations are insensitive to the absence the scalar degree of freedom.}
The above comment on the breaking of the degeneracy conditions and the scordatura mechanism applies when we investigate even-parity perturbations, which we leave for future study.

\subsection{Background field equations}
Now we assume a stealth solution with $X=X_0={\rm const.}$, which captures aspects of a broad range of known solutions.
For instance, a de Sitter attractor with constant $X$ was found in \cite{Armendariz-Picon:2000ulo}.
Also, there have been extensive studies on stealth S(dS) solutions~\cite{Mukohyama:2005rw,Babichev:2013cya,Kobayashi:2014eva,BenAchour:2018dap,Motohashi:2019sen,Takahashi:2019oxz,Minamitsuji:2019shy,Takahashi:2020hso} as well as stealth Kerr solutions~\cite{Charmousis:2019vnf,Takahashi:2020hso} in the context of Horndeski and (D)HOST theories.
\blue{As such, constant-$X$ solutions are motivated due to their link to exact stealth solutions while still having non-trivial scalar dynamics, therefore offering a controlled and simplified way to explore departures from GR. However, exact stealth solutions have also been found with non-constant $X$ in \cite{Bakopoulos:2023fmv} and their stability and effect on quasinormal modes have been quantified in \cite{Takahashi:2019oxz,Sirera:2024ghv}. Hence, looking forward, non-constant-$X$ solutions offer yet a richer and promising avenue to explore further viable GR departures.}
Under the condition that $X$ is a constant, one can express higher-derivative terms of $\phi$ as\footnote{At this point, we note that different notation conventions exist in the literature to rewrite combinations of derivatives acting on the scalar. In this work, we will not be using such conventions, but we refer to them here to facilitate the comparison of results. First, one can rewrite $\phi_{\alpha\beta}^2=\phi_\alpha^\beta\phi_\beta^\alpha$ and $\phi_{\alpha\beta}^3=\phi_\alpha^\beta\phi_\beta^\gamma\phi_\gamma^\alpha$ (more generally $\phi_{\alpha\beta}^n\equiv \phi_{\alpha_1}^{\alpha_2}\phi_{\alpha_2}^{\alpha_3}\cdots\phi_{\alpha_{n-1}}^{\alpha_n}\phi_{\alpha_n}^{\alpha_1}$), like it is done in e.g.~\cite{Takahashi:2020hso}. A second convention which was introduced in \cite{Zumalacarregui:2013pma} rewrites the same terms as $[\Phi^2]=\phi_\alpha^\beta\phi_\beta^\alpha$ and $[\Phi^3]=\phi_\alpha^\beta\phi_\beta^\gamma\phi_\gamma^\alpha$ (more generally $[\Phi^n]\equiv \phi_{\alpha_1}^{\alpha_2}\phi_{\alpha_2}^{\alpha_3}\cdots\phi_{\alpha_{n-1}}^{\alpha_n}\phi_{\alpha_n}^{\alpha_1}$).}
    \begin{align}
    \begin{split}
    &\phi^\mu\phi_\mn=0\;, \qquad
    \phi^\la\na_\mu\phi_{\nu\la}=-\phi_\mu^\la\phi_{\nu\la}\;, \qquad
    \phi^\la\Box\phi_\la=-\phi_\alpha^\beta\phi_\beta^\alpha\;,\\
    &\phi^\la\na_\la\phi_\mn=-R_{\mu\la\nu\si}\phi^\la\phi^\si-\phi_\mu^\la\phi_{\la\nu}\;, \qquad
    \phi^\la\na_\la\Box\phi=-R_{\la\si}\phi^\la\phi^\si-\phi_\alpha^\beta\phi_\beta^\alpha\;.
    \end{split}\label{eq:constantX}
    \end{align}
Then, the Euler-Lagrange (EL) equation for the metric is given by $\mE_\mn=0$, with
\begin{align}
\mE_\mn=& (F_2-X_0F_{3\phi}) G_{\mu\nu} \nonumber \\
    &-\frac{1}{2}\Bigl\{F_0+ 2X_0 (F_{1\phi} + 2 F_{2\phi\phi})+(A_1+A_2+2X_0B_{2\phi})\phi_\alpha^\beta\phi_\beta^\alpha -2\brb{F_{2\phi}+X_0(F_{3\phi\phi}-2A_{2\phi})}\Box\phi \nonumber \\
    &\quad +(F_{3\phi}-A_2)\brb{(\Box\phi)^2-\phi_\alpha^\beta\phi_\beta^\alpha- 2\phi^\alpha \phi^\beta R_{\alpha\beta}  }
    -2B_1\Box\phi\brb{(\Box\phi)^2-3\phi_\alpha^\beta\phi_\beta^\alpha- 3\phi^\alpha \phi^\beta R_{\alpha\beta}  }\nonumber \\
    &\quad  +6X_0B_{1\phi}(\Box\phi)^2+ (2B_2+B_3)\phi_\alpha^\beta\phi_\beta^\gamma\phi_\gamma^\alpha+2B_2\phi^\la\phi^\si\phi^{\alpha\beta}R_{\la\alpha\si\beta}\Bigl\}g_\mn \nonumber \\
    &-\frac{1}{2}\biggl\{F_{0X}+2(F_{1\phi}+F_{2\phi\phi})+ (F_{2X}-F_{3\phi})R + \left(F_{1X} - F_{3\phi\phi}-\frac{1}{2}RF_{3X}+ 4A_{2\phi} - 2X_0A_{3\phi} \right)\Box\phi \nonumber \\
    &\quad +\brb{A_{1X}+2(B_{2\phi}-X_0 B_{6\phi})}\phi_\alpha^\beta\phi_\beta^\alpha+\brb{A_{2X}+2(3B_{1\phi}-X_0 B_{4\phi})}(\Box\phi)^2+(B_{1X}+B_{4})(\Box\phi)^3\nonumber \\
    &\quad +F_{3X}\phi^{\alpha\beta}R_{\alpha\beta}+A_3\brb{(\Box\phi)^2-\phi_\alpha^\beta\phi_\beta^\alpha- \phi^\alpha \phi^\beta R_{\alpha\beta}  }+(B_{3X}-2B_6)\phi_\alpha^\beta\phi_\beta^\gamma\phi_\gamma^\alpha\nonumber\\
    &\quad +(B_{2X}-2B_4+B_6)\Box\phi\phi_\alpha^\beta\phi_\beta^\alpha-2B_4\Box\phi\phi^\al\phi^\beta R_{\al\beta}-2B_6\phi^\la\phi^\si\phi^{\alpha\beta}R_{\la\alpha\si\beta}\biggr\}\phi_\mu\phi_\nu \nonumber \\
    & -\Bigr\{F_{2\phi}+X_0(F_{3\phi\phi}+2A_{1\phi})-(F_{3\phi}+A_1-2X_0B_{2\phi})\Box\phi-B_2\brb{(\Box\phi)^2-\phi_\alpha^\beta\phi_\beta^\alpha-R_{\alpha\beta}\phi^\alpha\phi^\beta }\Bigr\}\phi_{\mu\nu}\nonumber \\
    & -\frac{1}{2}\big[2(F_{3\phi}+A_1+3X_0 B_{3\phi})+(2B_2-3B_3)\Box\phi\bigr]\phi_{\mu\la}\phi_\nu^\la -(2B_2+3B_3)\Box\phi_\la\phi_{(\mu}\phi_{\nu)}^\la \nonumber \\
    &-(F_{3\phi}+A_1+B_2\Box\phi)\phi^\la\phi^\si R_{\mu\la\nu\si} + 2B_2\phi^\la R_{\la\si}\phi_{(\mu}\phi_{\nu)}^\si-2(F_{3\phi}-A_2-3B_1\Box\phi)\phi^\la\phi_{(\mu}R_{\nu)\la}\nonumber \\
    &-2\brb{A_1+A_2+(3B_1+B_2)\Box\phi}\phi_{(\mu}\Box\phi_{\nu)} -3B_3\phi_{\la\si}\phi_\mu^\la\phi_\nu^\si -(2B_2+3B_3)\phi^{\la\si}\phi_{(\mu}\phi_{\nu)\la\si} \nonumber \\
    &+3B_3\phi^\la\phi^\si \phi_{(\mu}^\rho R_{\nu)\la\si\rho}+2B_2\phi^\la\phi^{\si\rho}\phi_{(\mu}R_{\nu)\si\la\rho}- T_{\mu\nu}\;,
    \label{EmnHOST}
\end{align}
where the coupling functions and their derivatives are evaluated at the background solution with $\phi=\phi_0$ (not necessarily a constant) and $X=X_0$.
Here, subscripts~$\phi$ and $X$ denote derivatives with respect to $\phi$ and $X$ respectively.  
The stress-energy tensor for the matter sector is given by $T_{\mu\nu} \equiv -\fr{2}{\sqrt{-g}} \fr{\delta(\sqrt{-g}\,L_{\rm m})}{\delta g^{\mu\nu}}$.
It should be noted that, unless $\phi={\rm const.}$, the EL equation for the scalar field, which we denote by $\mE_\phi=0$, is automatically satisfied for any configuration~$(g_\mn,\phi)$ that satisfies $\mE_\mn=0$ and the equations of motion for the matter field thanks to the Noether identity
associated with general covariance, i.e.~$\nabla^\mu{\cal E}_{\mu\nu}\propto \phi_\nu{\cal E}_\phi$
(see \cite{Motohashi:2016prk} for related discussions).
In other words, $\mE_\phi=0$ can be reproduced from other EL equations and hence is a redundant equation. Note also that the terms with $A_4$, $A_5$, $B_5$, $B_7$, $B_8$, $B_9$, and $B_{10}$ do not contribute to the EL equations under the condition~$X={\rm const}$. Because nothing about the metric background structure is assumed in Eq.~\eqref{EmnHOST}, this is our main equation that we will use to derive the existence conditions for general GR metrics, either with minimally coupled matter field(s) or in vacuum.

Equation~\eqref{EmnHOST} is the master field equation derived in this work. From it, by taking different limits, one can recover the relevant field equations for the different cases collected in Table~\ref{main-table}. More concretely, one can easily restrict Eq.~\eqref{EmnHOST} to shift-symmetric cubic HOST by rewriting all the functions as $F(\phi,X)=F(X)$, and therefore setting their $\phi$ derivatives to zero, i.e.~$F_\phi=F_{\phi\phi}=0$. It is also straightforward to restrict to quadratic HOST theories by setting the cubic functions to zero, i.e.~$B_J=F_3=0$. In the latter case, Eq.~(\ref{EmnHOST}) recovers the expressions derived in \cite{Takahashi:2020hso}.

Let us see now how the equation of motion gets simplified as we progressively weaken the requirement on background solutions and only require the presence of subsets of the previous solutions. 
Assuming that $G_\mn=-\Lambda g_\mn$, Eq.~(\ref{EmnHOST}) reduces to
\begin{align}
    \mE_\mn=
        &-\frac{1}{2}\biggl\{F_0+2\Lambda F_2+2X_0\left[F_{1\phi}+2F_{2\phi\phi}+2\Lambda\left(\frac{1}{2}F_{3\phi}-A_1-2A_2+6X_0 B_{1\phi}\right)\right]\nonumber\\
        &\quad -2\left[F_{2\phi\phi}+X_0(F_{3\phi\phi}-2A_{2\phi}+2\Lambda(5B_1+2B_2))\right]\Box\phi+B_3\phi_\alpha^\beta\phi_\beta^\gamma\phi_\gamma^\alpha\nonumber\\
        &\quad +\left[A_1+A_2+2X_0(3B_{1\phi}+B_{2\phi})+2(2B_1+B_2)\Box\phi\right]\phi_\alpha^\beta\phi_\beta^\alpha\biggl\}g_\mn\nonumber\\
        &-\frac{1}{2}\bigl\{F_{0X}+2F_{1\phi}+2F_{2\phi\phi}+2\Lambda[2F_{2X}-F_{3\phi}-A_1-2A_2+X_0(2A_{2X}+3A_3+4(3B_{1\phi}-X_0B_{4\phi}))]\nonumber\\
        &\quad +\left[F_{1X}-F_{3\phi\phi}+4A_{2\phi}-2X_0A_{3\phi}-\Lambda(F_{3X}+12B_1+2B_2-4X_0(B_{1X}+2B_4+B_6))\right]\Box\phi\nonumber\\
        &\quad+\left[A_{1X}+A_{2X}+2(3B_{1\phi}+B_{2\phi}-X_0(B_{4\phi}+B_{6\phi}))+(B_{1X}+B_{2X}-B_4-B_6)\Box\phi\right]\phi_\alpha^\beta\phi_\beta^\alpha\nonumber\\
        &\quad+B_{3X}\phi_\alpha^\beta\phi_\beta^\gamma\phi_\gamma^\alpha\bigl\}\phi_\mu\phi_\nu\nonumber\\
        &-\bigl[F_{2\phi}+X_0(F_{3\phi\phi}+2A_{1\phi}-2\Lambda B_2)+B_2\phi_\alpha^\beta\phi_\beta^\alpha+2X_0B_{2\phi}\Box\phi\bigl]\phi_{\mn}-3\left(X_0B_{3\phi}-\frac{1}{2}B_3\Box\phi\right)\phi_{\mu\lambda}\phi^\lambda_\nu\nonumber\\
        &-2\bigl[A_1+A_2+(3B_1+B_2)\Box\phi\bigl]\phi_{(\mu}\Box\phi_{\nu)}-3B_3\left(\phi_{\lambda\sigma}\phi^\lambda_\mu\phi^\sigma_\nu+\phi_{(\mu}\phi_{\nu)\lambda\sigma}\phi^{\lambda\sigma}-\phi^\lambda\phi^\sigma\phi_{(\mu}^\rho R_{\nu)\lambda\sigma\rho}\right)\nonumber\\
        &-B_2\left(\phi_{(\mu}\phi_{\nu)\lambda\sigma}\phi^{\lambda\sigma}-\phi^\lambda\phi^{\sigma\rho}\phi_{(\mu}R_{\nu)\sigma\lambda\rho}\right)-\left(2B_2+3B_3\right)\Box\phi_\lambda\phi_{(\mu}\phi^\lambda_{\nu)}\;,
        \label{EmnHOSTSdS}
\end{align}
where we have used the identities
 \begin{align}
    (\Box\phi)\phi_\mn- \phi_\mu^\la\phi_{\la\nu}- R_{\mu\la\nu\si}\phi^\la\phi^\si
    =\Lambda\left(2X_0 g_\mn +\phi_\mu\phi_\nu\right)\;, \qquad
    (\Box\phi)^2-\phi_\alpha^\beta\phi_\beta^\alpha&=4\Lambda X_0 \;.\label{id}
    \end{align}
Notice that the above identities hold true in the case of the S(dS) spacetime with a linearly time-dependent scalar field~$\phi=q t+\psi(r)$,\footnote{\blue{\blue{The coordinates~$t$ and $r$ here} are static and spherically symmetric coordinates as defined in Eq.~\eqref{eq:bg_metric_sp}.}} so that $X_0=q^2/2$. Here $q$ has a mass dimension two for a scalar with mass dimension one.\footnote{Note that sometimes $\mu$ is used instead of $q$ to denote the time-dependence in the scalar.}
Having more simplified equations of motion then leads to less restrictive existence conditions, and hence the case $\textbf{Cubic}_{\text{SdS}}$, where only SdS solutions are required to exist, will generally allow a bigger region of theory space than in case $\textbf{Cubic}_{\text{GR-mat}}$. 
In the limit of including only shift-symmetric quadratic terms (i.e.~$\prescript{\text{SS}}{}{\textbf{Quadratic}}_{\text{SdS}}$), Eq.~\eqref{EmnHOSTSdS} above reduces to Eq.~(2.9) in \cite{Takahashi:2020hso}.

We can further weaken the requirement on the background geometry to be that of a Schwarzschild black hole, i.e.~if we are interested in vacuum solutions where $G_\mn=0$ (hence with $\Lambda=0$), then Eq.~(\ref{EmnHOSTSdS}) is further simplified to
\begin{align}
    \mE_\mn=
        &-\frac{1}{2}\biggl\{F_0+2X_0\left(F_{1\phi}+2F_{2\phi\phi}\right)-2\left[F_{2\phi}+X_0(F_{3\phi\phi}-2A_{2\phi})\right]\Box\phi\nonumber\\
        &\qquad +\left[A_1+A_2+2X_0(3B_{1\phi}+B_{2\phi})+2\left(2B_1+B_2+\frac{11}{18}B_3\right)\Box\phi\right]\phi_\alpha^\beta\phi_\beta^\alpha\biggl\}g_\mn\nonumber\\
        &-\frac{1}{2}\biggl\{F_{0X}+2F_{1\phi}+2F_{2\phi\phi}+\left(F_{1X}-F_{3\phi\phi}+4A_{2\phi}-2X_0A_{3\phi}\right)\Box\phi\nonumber\\
        &\qquad+\bigg[A_{1X}+A_{2X}+2(3B_{1\phi}+B_{2\phi}-X_0(B_{4\phi}+B_{6\phi}))\nonumber\\
        &\qquad\qquad+\left(B_{1X}+B_{2X}+\frac{5}{9}B_{3X}-\frac{1}{3}B_3-B_4-B_6\right)\Box\phi\bigg]\phi_\alpha^\beta\phi_\beta^\alpha\biggl\}\phi_\mu\phi_\nu\nonumber\\
        &-\brb{F_{2\phi}+X_0(F_{3\phi\phi}+2A_{1\phi})+\frac{1}{2}(2B_2+B_3)\phi_\alpha^\beta\phi_\beta^\alpha+2X_0B_{2\phi}\Box\phi}\phi_{\mn}-3X_0B_{3\phi}\Box\phi\phi_{\mu\lambda}\phi^\lambda_\nu\nonumber\\
        &-2\biggl[A_1+A_2+\left(3B_1+2B_2+\frac{7}{6}B_3\right)\Box\phi\biggr]\phi_{(\mu}\Box\phi_{\nu)}-(2B_2+B_3)\Box\phi_\lambda\phi_{(\mu}\phi^\lambda_{\nu)}\;,
        \label{EmnHOSTSchw}
\end{align}
where we have used the following identities:
   \begin{align}
    \begin{split}
    &\phi_\alpha^\beta\phi_\beta^\gamma\phi_\gamma^\alpha = \frac{5}{9} \phi_\alpha^\beta\phi_\beta^\alpha \Box \phi \;, \qquad
    X_0\phi^{\sigma\lambda}\phi_{\mu\sigma\lambda} = \left[X_0\Box\phi_\mu
    -\frac{2}{9}(\Box\phi)^2\phi_\mu\right]\Box\phi \;, \\
    &X_0 R_{\mu\sigma\lambda\rho} \phi^{\sigma\rho} \phi^\lambda = -\frac{2}{5} \phi_\alpha^\beta\phi_\beta^\gamma\phi_\gamma^\alpha \phi_\mu\;, \qquad
    R_{\mu\sigma\lambda\rho} \phi^\sigma \phi^\lambda \phi^\rho_\nu = \phi^\sigma_\mu (-\Box\phi \phi_{\nu\sigma} + \phi_\nu^\rho \phi_{\sigma\rho})\;, \\
    &X_0\brb{9\Box\phi\phi_\mu^\sigma\phi_{\nu\sigma}+4\phi_{(\mu}\left(\Box\phi\Box\phi_{\nu)}+3\phi_{\nu)}^\sigma\Box\phi_{\sigma}\right)}=(\Box\phi)^2\brb{3X_0\phi_{\mu\nu}+\Box\phi\left(2X_0 g_{\mu\nu}+3\phi_\mu\phi_\nu\right)}\;,\\
    \end{split}
    \end{align}
which are valid for the stealth Schwarzschild background.
Indeed, Eq.~\eqref{EmnHOSTSchw} can be used to derive the existence conditions for Schwarzschild black holes.
We will see that weakening of the nature of the background allows more freedom in the remaining valid theory space, and hence also in the potential number of GR deviations, as can be seen in Table~\ref{main-table} and Figure~\ref{fig:cH-functions-Schw}.

Finally, for completeness, let us also comment on the trivial case where $\phi={\rm const.}$, and hence all derivatives of $\phi$ vanish.
The equations of motion for the metric and scalar are given by
\begin{align}
    \begin{split}
    \mE_\mn&= F_2 G_{\mu\nu} -\frac{1}{2}F_0g_\mn-T_\mn=0 \;,\\
    \mE_\phi&= F_{0\phi}+F_{2\phi}R=0 \;,
    \end{split}
\end{align}
which coincide with the expressions in \cite{Takahashi:2020hso} for quadratic HOST.
Note that the equation of motion of $\phi$ cannot be derived from that of the metric in this case.
In what follows, we only consider the case where the scalar field has a non-trivial gradient and therefore ${\cal E}_\phi=0$ automatically follows from ${\cal E}_{\mu\nu}=0$.

\subsection{Existence conditions for stealth GR solutions}
We can now obtain the conditions that need to be satisfied for each class of stealth solutions to exist. We will show here the conditions for general cubic theories (i.e.~cases $\textbf{Cubic}_{\text{GR-mat/GR-vac/SdS/Schw}}$), while an exhaustive list of existence conditions for shift-symmetric and/or quadratic theories can be found in Appendix~\ref{app-existconds}.

Note that we obtain the existence conditions for stealth solutions in such a way that the covariant equations of motion are trivially satisfied when the Einstein equation in GR (i.e.~$G_{\mu\nu} = \Mpl^{-2} T_{\mu\nu} - \Lambda g_{\mu\nu}$) is imposed.
Once we assume such existence conditions, the metric is determined by solving the Einstein equation under the spacetime symmetry of interest, and then the scalar field profile is fixed via $X_0=-g^{\mu\nu}\partial_\mu\phi_0\partial_\nu\phi_0/2$~\cite{Charmousis:2019vnf,Takahashi:2020hso}.
It should also be noted that the existence conditions are written in terms of the functions of HOST theories evaluated at the background solution with $\phi=\phi_0$ and $X=X_0$.
Since we are focusing on solutions with $X_0={\rm const.}$, even if the existence conditions impose, e.g.~$A_1(\phi_0,X_0)=0$, this does not necessarily mean $A_{1X}(\phi_0,X_0)=0$.
However, the condition~$A_1(\phi_0,X_0)=0$ implies $A_{1\phi}(\phi_0,X_0)=0$ when $\phi_0$ is a non-trivial function of spacetime.
(Recall that we do not impose the shift symmetry from the outset, and therefore the functions can depend on $\phi$ explicitly.)
This property has been used to simplify the conditions.

\subsubsection{General stealth GR with minimally coupled matter}
Let us begin by investigating the case~$\textbf{Cubic}_{\text{GR-mat}}$, in which Eq.~\eqref{EmnHOST} is required to allow general GR solutions (i.e.~which satisfy $G_{\mu\nu} = \Mpl^{-2} T_{\mu\nu} - \Lambda g_{\mu\nu}$).
If the following conditions,
\begin{equation}\label{eq:existcondGRT}
    \begin{aligned}
    &F_0+2\Lambda\Mpl^2=-2X_0(F_{1\phi}+2F_{2\phi\phi})=X_0(F_{0X}-2F_{2\phi\phi}) \;,\\
    &X_0F_{1X}=-3F_{2\phi}=-3X_0F_{3\phi\phi}\;,\quad X_0^{-1}(F_2-\Mpl^2)=F_{2X}=F_{3\phi}=-A_1=A_2 \;,\\
    &F_{3X}=A_{1X}=A_{2X}=A_3=B_1=B_{1X}=B_2=B_{2X}=B_3=B_{3X}=B_4=B_6=0 \;,
\end{aligned}
\end{equation}
are satisfied at $X=X_0$,\footnote{In writing the conditions %below
in \eqref{eq:existcondGRT}, we have also employed the relation~$R=4\Lambda-\Mpl^{-2}T$, where $T\equiv T_\mu^\mu$ is the trace of the stress-energy tensor.}
this \blue{sufficiently} ensures that Eq.~\eqref{EmnHOST} is satisfied for any stealth GR solution with arbitrary time-dependent scalar background with constant $X_0$, and any minimally coupled matter field---this is visualised in Figure~\ref{fig:cH-functions-generalGR}.\footnote{\blue{Note that the specific colour chosen for the boxes in Figure~\ref{fig:cH-functions-generalGR} is not important. Instead, the number of different colours indicates the number of independent combinations of HOST functions after the imposition of existence conditions. In the present case, we see that after the imposition of the conditions~\eqref{eq:existcondGRT}, all non-zero HOST functions can be rewritten in terms of four independent functions, e.g.~$\{F_0,F_{1\phi},F_2,F_{2\phi}\}$.}}
In that sense, they are the most restrictive set of existence conditions presented in this paper.
The conditions~\eqref{eq:existcondsGRmatSSCub} for $\prescript{\text{SS}}{}{\textbf{Cubic}}_{\text{GR-mat}}$ can be obtained from Eq.~\eqref{eq:existcondGRT} by imposing shift symmetry on all the functions. Additionally, the conditions~\eqref{eq:existcondsGRmatQuad} for $\textbf{Quadratic}_{\text{GR-mat}}$ can be obtained by removing all cubic-order interactions, and then similarly the conditions~\eqref{eq:existcondsGRmatSSQuad} for $\prescript{\text{SS}}{}{\textbf{Quadratic}}_{\text{Gr-mat}}$ can be obtained by then taking the shift-symmetric limit, recovering the expressions in \cite{Takahashi:2020hso}.

\subsubsection{General stealth GR in vacuum}\label{sec:genGRstealth-vac}
In the case of stealth GR metric solutions in vacuum, i.e.~if we focus on $\textbf{Cubic}_{\text{GR-vac}}$, the existence conditions for cubic HOST theories now read
\begin{align}
    &F_0+2\Lambda(F_2-X_0F_{3\phi})=-2X_0(F_{1\phi}+2F_{2\phi\phi})\;,\quad F_{0X}=-2[F_{1\phi}+F_{2\phi\phi}+\Lambda(2F_{2X}-2F_{3\phi}+X_0A_{1X})]\;,\nonumber\\
    &3F_{2\phi}+X_0(F_{1X}-\Lambda F_{3X})=2X_0^2(A_{3\phi}-2\Lambda B_4)\;,\quad F_{2\phi}=X_0F_{3\phi\phi}\;,\quad F_{3\phi}=-A_1=A_2\;,\nonumber\\
    &A_3=A_{1X}=-A_{2X}+2X_0B_{4\phi}\;,\quad 2B_4=-2B_{1X}=B_{2X}\;,\quad  B_1=B_2=B_3=B_{3X}=B_6=0\;.
    \label{eq:existcondGRV}
\end{align}
From the above conditions, we see that less functions are required to be set to zero, therefore allowing a bigger region in the theory parameter space.
Similarly as before, the conditions for $\prescript{\text{SS}}{}{\textbf{Cubic}}_{\text{GR-vac}}$~\eqref{eq:existcondsGRvacSSCub}, $\textbf{Quadratic}_{\text{GR-vac}}$~\eqref{eq:existcondsGRvacQuad}, and $\prescript{\text{SS}}{}{\textbf{Quadratic}}_{\text{GR-vac}}$~\eqref{eq:existcondsGRvacSSQuad} can be obtained from Eq.~\eqref{eq:existcondGRV} in the appropriate limits. The results we have obtained for $\textbf{Quadratic}_{\text{GR-vac}}$ and $\prescript{\text{SS}}{}{\textbf{Quadratic}}_{\text{GR-vac}}$ coincide with those found in \cite{Takahashi:2020hso}.

\subsubsection{Schwarzschild-de Sitter}\label{sec:SdS}
Here we focus on extracting the existence conditions for SdS background solutions in cubic HOST theories, i.e.~$\textbf{Cubic}_{\text{SdS}}$. From Eq.~\eqref{EmnHOSTSdS} we obtain
\begin{align}
&F_0+2\Lambda F_2=-2X_0(F_{1\phi}+2F_{2\phi\phi}+\Lambda F_{3\phi}+2\Lambda A_{1})\;, \quad F_{2\phi}=-X_0(F_{3\phi\phi}-2A_{2\phi})\;,\nonumber \\
&F_{0X}+4\Lambda F_{2X}=-2\{F_{1\phi}+F_{2\phi\phi}-\Lambda[F_{3\phi}-A_1-X_0(2A_{2X}+3A_3-4X_0 B_{4\phi})]\}\;,\quad A_{1}=-A_{2}\;,\nonumber \\
&A_{1X}+A_{2X}=2X_0(B_{4\phi}+B_{6\phi})\;,\quad F_{1X}-\Lambda F_{3X}=F_{3\phi\phi}-4A_{2\phi}+2X_0 A_{3\phi}-4\Lambda X_0(2B_4+B_6+B_{1X})\;,\nonumber\\
&B_1=B_2=B_3=B_{3X}=0\;,\quad B_4+B_6=B_{1X}+B_{2X}\;.
    \label{eq:existcondSdS}
\end{align}
These conditions are displayed in Figure~\ref{fig:cH-functions-SdS}. 
When shift-symmetry is imposed, i.e.~$\prescript{\text{SS}}{}{\textbf{Cubic}}_{\text{SdS}}$, the above conditions reduce to those derived in \cite{Minamitsuji:2019shy}.
We can also obtain the conditions up to quadratic interactions, i.e.~$\textbf{Quadratic}_{\text{SdS}}$, from the equation above by setting all $B_J=0$ as well as $F_3=0$ [see \eqref{eq:existcondsSdSQuad}]. Finally, in the case of shift-symmetric quadratic theories, i.e.~$\prescript{\text{SS}}{}{\textbf{Quadratic}}_{\text{SdS}}$, Eq.~\eqref{eq:existcondSdS} recovers Eq.~(42) of \cite{Motohashi:2019sen} [see \eqref{eq:existcondsSdSSSQuad}].

\subsubsection{Schwarzschild}\label{sec:Schw}
\begin{figure}
\centering
\begin{subfigure}{0.95\textwidth}
    \includegraphics[width = 0.95\textwidth]{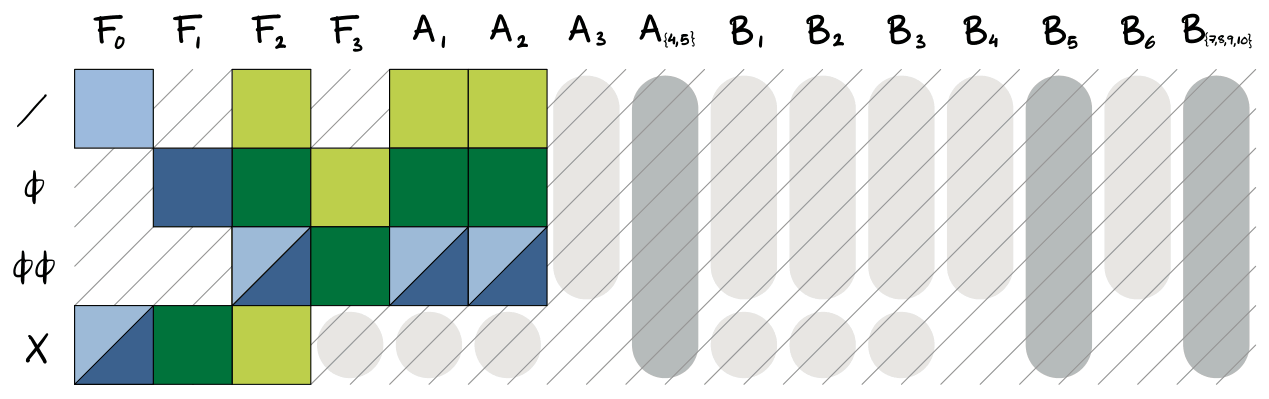}
    \caption{Existence conditions for general GR stealth solutions with matter ($\textbf{Cubic}_{\text{GR-mat}}$) \eqref{eq:existcondGRT}.}
    \label{fig:cH-functions-generalGR}
\end{subfigure}
\hfill
\begin{subfigure}{0.95\textwidth}
    \includegraphics[width = 0.95\textwidth]{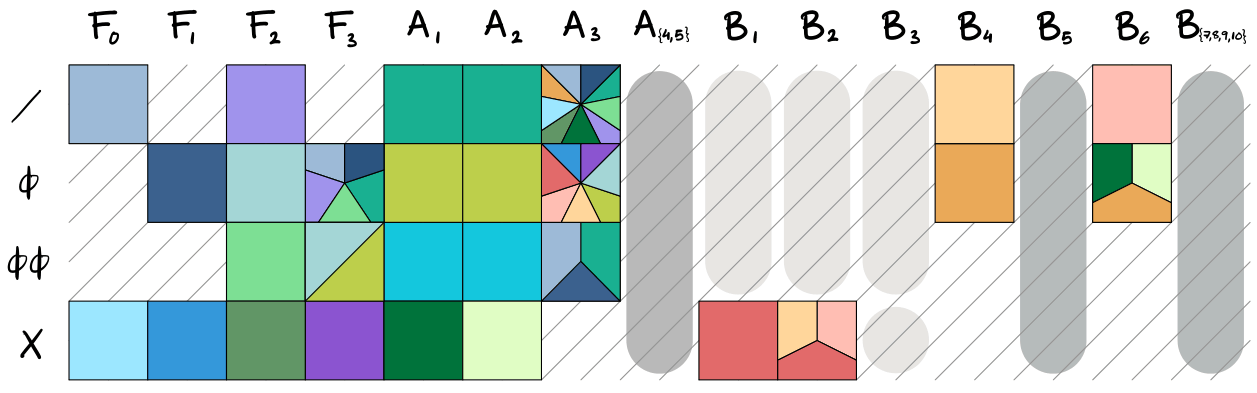}\\
    \caption{Existence conditions for stealth %Schwarzschild-de Sitter
    SdS solutions ($\textbf{Cubic}_{\text{SdS}}$) \eqref{eq:existcondSdS}.}
    \label{fig:cH-functions-SdS}
\end{subfigure}
\hfill
\begin{subfigure}{0.95\textwidth}
    \includegraphics[width = 0.95\textwidth]{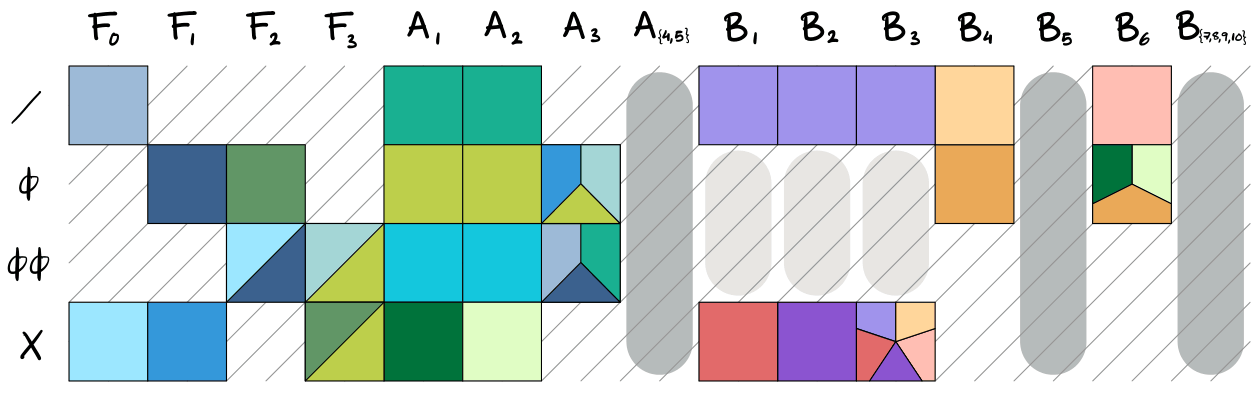}\\
    \caption{Existence conditions for stealth Schwarzschild solutions ($\textbf{Cubic}_{\text{Schw}}$) \eqref{eq:existcondSchw}.}
    \label{fig:cH-functions-Schw}
\end{subfigure}
\caption{Here we display the theory space of cubic HOST theories which allow the existence of different stealth solutions. (Rows) columns represent (derivatives of) cubic HOST functions, e.g.~$\phi\phi$-row and $F_0$-column refers to $F_{0\phi\phi}$.
%In coloured boxes, we show all the functions that are allowed to play a role in each case, with different colours indicating independent (combinations of) free functions after the imposition of the corresponding conditions.
%When a single box is split into regions of different colors—as in the case of 
%in subfigure (b)—this indicates that distinct, independent combinations of functions contribute in that term. We will revise the figure caption to make this explanation more explicit and improve clarity."
\blue{
%In coloured boxes, we show contributing functions. Colours indicate (combinations of) free functions that remain independent after the imposition of the corresponding existence conditions. When a single box is split into regions of different colours, this indicates that such term can be rewritten as a distinct combination of functions corresponding to such colours.
In coloured boxes, we show (combinations of) free functions that remain independent after the imposition of the corresponding existence conditions. When a single box is split into regions of different colours, this indicates that such term can be rewritten as a distinct combination of functions corresponding to such colours.
}
The shaded region is either not allowed due to the %$X_0=\text{const.}$
constant-$X$ nature of our background (dark grey), or due to the existence conditions (light grey). Finally, uncoloured and unshaded regions simply do not contribute to the equations of motion. Note that the allowed theory-space region increases as the assumptions on the background geometry are weakened.
}
\label{fig:figures}
\end{figure}

Here we consider the conditions that HOST functions are required to satisfy in order to admit the existence of stealth Schwarzschild solutions ($\textbf{Cubic}_{\text{Schw}}$).
From Eq.~(\ref{EmnHOSTSchw}) we obtain
\begin{align} 
&F_0=-2X_0(F_{1\phi}+2F_{2\phi\phi})\;, \quad F_{2\phi}=-X_0(F_{3\phi\phi}-2A_{2\phi})\;,\quad F_{0X}=-2(F_{1\phi}+F_{2\phi\phi})\;,\quad A_{1}=-A_{2}\;,\nonumber \\
&A_{1X}+A_{2X}=2X_0(B_{4\phi}+B_{6\phi})\;,\quad F_{1X}=F_{3\phi\phi}-4A_{2\phi}+2X_0 A_{3\phi}\;,\quad 18B_1=2B_2=-B_3\;,\nonumber\\
&3B_3=X_0[9(B_{1X}+B_{2X}-B_4-B_6)+5B_{3X}]\;,\quad B_{1\phi}=B_{2\phi}=B_{3\phi}=0\;.
    \label{eq:existcondSchw}
\end{align}
These conditions are shown in Figure~\ref{fig:cH-functions-Schw}. \blue{In this case, it is important to clarify why certain functions---such as $F_2$---appear uncoloured and unshaded in Figure~\ref{fig:cH-functions-Schw}, in contrast to their coloured or shaded appearance in Figures~\ref{fig:cH-functions-generalGR} and \ref{fig:cH-functions-SdS}. Taking $F_2$ as an example, we note that in a Ricci-flat geometry like Schwarzschild, its contribution becomes redundant, since $F_2G_{\mu\nu}=0$ automatically holds in Eq.~\eqref{EmnHOST}. This does not imply that $F_2$ is inadmissible in this case; rather, it means that $F_2$ is unconstrained by the imposition of the existence of a Schwarzschild background and is therefore allowed complete freedom.} Similarly to the previous case, in the shift-symmetric limit, corresponding to $\prescript{\text{SS}}{}{\textbf{Cubic}}_{\text{Schw}}$, these conditions exactly reproduce the results in \cite{Minamitsuji:2019shy} and given in \eqref{eq:existcondsSchwSSCub}. For quadratic theories, i.e.~$\textbf{Quadratic}_{\text{Schw}}$, the existence conditions can also be obtained from the equation above by setting all $B_J=0$ as well as $F_3=0$ [see \eqref{eq:existcondsSchwQuad}]. Finally, in the case of shift-symmetric quadratic theories, i.e.~$\prescript{\text{SS}}{}{\textbf{Quadratic}}_{\text{Schw}}$, Eq.~\eqref{eq:existcondSdS} recovers the results of Eq.~(23) in \cite{Motohashi:2019sen} and given in \eqref{eq:existcondsSchwSSQuad}.

At this point, it is interesting to inspect the existence conditions obtained so far, specifically in regards to the degeneracy conditions summarised in section~\ref{sec:HOST}.
One notices that the existence conditions~\eqref{eq:existcondGRT}, \eqref{eq:existcondGRV}, and \eqref{eq:existcondSdS} impose $B_1=0$ when evaluated at the background.
Also, the existence condition~\eqref{eq:existcondSchw} for the stealth Schwarzschild solution imposes $18B_1=2B_2=-B_3$, which implies $B_1=0$ when the degeneracy condition~\eqref{DC_cubic} for cubic DHOST of class~${}^3$N-I is assumed.
Recall that for cubic DHOST of class~${}^3$N-I, one requires that $B_1\neq0$ [see \eqref{DC_cubic} and \eqref{DC_quadratic+cubic}, where $B_1$ appears in the denominator].
This means that, if we work within DHOST theories of class N-I and require the existence of stealth solutions with constant %$X_0$
$X$, the cubic DHOST part is prohibited, and we are left with quadratic DHOST of class ${}^2$N-I (see also \cite{Minamitsuji:2019shy}).\footnote{Here, we have assumed that the theory is valid all the way from the black hole scale to the cosmological scale. Recall that, as mentioned in section~\ref{sec:HOST}, the class~N-I (i.e., a sum of quadratic DHOST of class~${}^2$N-I and cubic DHOST of class~${}^3$N-I) is the only subclass of DHOST that allows for viable cosmology.}
Regarding the degeneracy condition~\eqref{DC_quadratic} for quadratic DHOST, one easily sees the compatibility with the existence conditions.
Indeed, all the existence conditions~\eqref{eq:existcondGRT}--\eqref{eq:existcondSchw} satisfy $A_2=-A_1$, and the functions~$A_4$ and $A_5$ are irrelevant to the existence conditions.
In the remaining of the paper, however, and as motivated in the introduction, we will not be requiring the satisfaction of degeneracy conditions.

\section{Quadratic Lagrangian in the odd sector}\label{sec:perturbations}

In the previous section we have investigated the background evolution in cubic HOST theories as well as in a number of specific subcases, with a particular focus on understanding the constraints imposed by requiring stealth GR solutions. Having done so, we arrived at a reduced set of cubic HOST theories and in this section we now consider perturbations about the stealth solutions to ultimately understand how restrictions imposed by requiring GR stealth solutions affect the behaviour of perturbations (and ultimately observable quasinormal modes). Specifically, we here investigate the dynamics of linear odd-parity perturbations about a static and spherically symmetric background given by
\begin{equation}\label{eq:bg_metric_sp}
    {\rm d}s^2=-A(r) \de t^2 + \frac{\de r^2}{B(r)} + r^2 \left(\de {\theta}^2 + {\sin}^2\theta\,%\de {\phi}^2
    {\rm d}\varphi^2\right)\;,
\end{equation}
accompanied by a spherical and (linearly) time-dependent scalar field~$\phi=q t+\psi(r)$.
In this section, we will focus on S(dS) solutions, i.e.~the unique static and spherically symmetric vacuum solutions of GR, therefore satisfying $A=B$. We, however, choose to keep $A$ and $B$ independent for now, as the equations we derive can also be applied to study the dynamics of perturbations for non-stealth metrics with $A\neq B$, e.g.~those corresponding to hairy black holes~\cite{Babichev:2013cya,Kobayashi:2014eva,Babichev:2016fbg}. \blue{We begin by studying the dynamics of odd-parity perturbations as an initial step toward identifying stable models within the broad spectrum of theories explored in this work, reserving a more comprehensive analysis of the more complex even-parity modes for future investigation.\footnote{\blue{The additional complexity arises from the higher number of even-parity functions as well as their coupling to the scalar mode.}}}

In the so-called Lema\^{i}tre coordinates, the line element~\eqref{eq:bg_metric_sp} can be written as:
\begin{align}\label{eq:bg_metric_Le}
     {\rm d}s^2=-\de \tau^2 + (1-A) \de \rho^2 + r^2 \left(\de {\theta}^2 + {\sin}^2\theta\,
     {\rm d}\varphi^2\right)\;,
\end{align}
where $\tau$ and $\rho$ are defined so that
\begin{align}\label{eq:coord-transf}
    \de \tau = \de t + \sqrt{\frac{1-A}{AB}}~\de r \;, \qquad \de \rho = \de t + \frac{\de r}{\sqrt{AB(1-A)}} \;.
\end{align}
We then see that the coordinate~$r$ is a function of $\rho -\tau$, satisfying
\begin{align}
    \partial_\rho r = -\dot{r} = \sqrt{\frac{B(1-A)}{A}} \;, \label{drdrho-drdtau}
\end{align}
where a dot denotes the derivative with respect to $\tau$. As a reminder, recall that in the previous sections we have distinguished four different cases in relation to the nature of the background solution: 1)~general GR solutions in the presence of matter, 2)~general GR solutions in vacuum, 3)~SdS, and 4)~Schwarzschild black holes. As a result, for cases~1) and 2) one can in principle study the dynamics of perturbations for metrics other than S(dS), e.g.~Kerr for cases~1) and 2) and Reissner-Nordstr{\"o}m for case~1). However, in this section we will focus on static and spherically symmetric stealth GR solutions in vacuum, i.e., stealth S(dS) solutions.
Written explicitly, the metric functions for these solutions are as follows:
\begin{align}
    \text{Schwarzschild:}&\hspace{20pt} A=B=1-\frac{r_s}{r}\;, \label{eq:Schw}\\
    \text{SdS:}&\hspace{20pt} A=B=1-\frac{r_s}{r}-\frac{1}{3}\Lambda r^2\;,\label{eq:SdS}
\end{align}
where $r_s\,(>0)$ is a constant of length dimension (corresponding to the horizon radius for the Schwarzschild metric). Note that for the S(dS) metric, the following relation applies:
\be
    A'-\frac{1-A}{r}=-\Lambda r \;.
    \label{eq:SdSA}
\ee
We now introduce metric perturbations~$h_{\mu\nu}$ as
\begin{align}
    h_{\mu\nu} &\equiv g_{\mu\nu} - \bar g_{\mu\nu} \;,
\end{align}
where $\bar g_{\mu\nu}$ is the background metric given by Eq.~\eqref{eq:bg_metric_Le}.
As we are considering the odd-parity sector and hence no scalar perturbations will be present in our analysis, we will (in an abuse of notation) use the same symbol for the scalar field~$\phi$ and its background value, i.e.~$\phi_0=\phi$. Note that, if we assume that the Lema\^{i}tre coordinate~$\tau$ is compatible with the time coordinate in the unitary gauge (i.e., $\phi\propto \tau$), then $X_0$ is a constant.

The odd-parity metric perturbations are usually decomposed in terms of the spherical harmonics~$Y_{\ell m}(\theta,\varphi)$.
Note that one can set $m=0$ without loss of generality thanks to the spherical symmetry of the background,\footnote{\blue{This is because in spherically symmetric backgrounds all $m$-modes are degenerate in frequency and decouple from each other at linear level.}} and hence one can employ the Legendre polynomials~$P_\ell(\cos\theta)$ instead.
In the Regge-Wheeler gauge, the odd-parity metric perturbations look like
\begin{align}
    h_{\mu\nu}^{\rm odd}=\left(\begin{array}{cccc}
        0 & 0 & 0 & h_0\\
        0 & 0 & 0 & h_1\\
        0 & 0 & 0 & 0\\            
        h_0 & h_1 & 0 & 0
    \end{array} \right)r^2\sin{\theta}\,\partial_\theta %Y_{\ell 0}
    P_\ell(\cos\theta)\;,
    \label{eq:hodd}
\end{align}
where the $\ell$-dependence of $h_0$ and $h_1$ has been suppressed as modes with different $\ell$ evolve independently.
Note in passing that the Regge-Wheeler gauge can be achieved by a complete gauge fixing, and therefore one can impose it at the level of Lagrangian~\cite{Motohashi:2016prk}.
Here, $h_0$ and $h_1$ are functions of $(\tau,\rho)$, and it should be pointed out that, as can be seen above, they differ from definitions used in other works (e.g.~\cite{Langlois:2021aji,Tomikawa:2021pca,Takahashi:2019oxz,Sirera:2023pbs}) by an overall factor of $r^2$ which has been included for later convenience.
It should also be noted that we focus on generic higher multipoles with $\ell\ge 2$, where one expects to have one propagating degree of freedom in the odd sector.
For $\ell=1$, the odd-parity perturbations are non-dynamical and correspond to a slow rotation of the black hole.

We write the perturbed covariant action up to quadratic order in metric perturbations as:
\begin{align}
    S_{\rm grav}^{(2)}=\frac{1}{4}\int {\rm d}^4x\sqrt{-g}\sum_a\biggl[
    &\sum_{K=0}^{3}\delta\mathcal{L}_{F_{Ka}}F_{Ka}
    +\sum_{I=1}^5\delta\mathcal{L}_{A_{Ia}}A_{Ia}+\sum_{J=1}^{10}\delta\mathcal{L}_{B_{Ja}}B_{Ja}\biggr] \;, \label{eq:quadaction1}
\end{align}
where we have introduced the new notation~$a=\{\emptyset,\phi,X,\phi\phi,XX,\phi X\}$ denoting the different $\phi$- and $X$-derivatives, with $\emptyset$ referring to a lack of derivatives. This notation enables one to easily identify which terms contribute at perturbative level and also directly confirm which features of the background solution cause the other terms to vanish. For the setup considered here with linear odd-parity perturbations with $X={\rm const.}$, out of the potential 114 contributions to \eqref{eq:quadaction1}, only 23 are non-zero, which we collect in Appendix~\ref{app-covquadL}.
As summarised in Figure~\ref{fig:cubic-HOST-functions-quad-L}, the rest of the terms vanish upon employing 
constant-$X$ identities
Eq.~\eqref{eq:constantX} and/or other simplifying relations that hold true for odd-parity perturbations about spherical background (such as $h^\mu_\mu=0$ or $\phi^\mu\phi^\nu h_{\mu\nu}=0$) which are collected in full in Eq.~\eqref{eq:odd-simp-rels} of Appendix~\ref{app-covquadL}.

\begin{figure}
    \centering
    \centering
    \includegraphics[width = 0.99\textwidth]{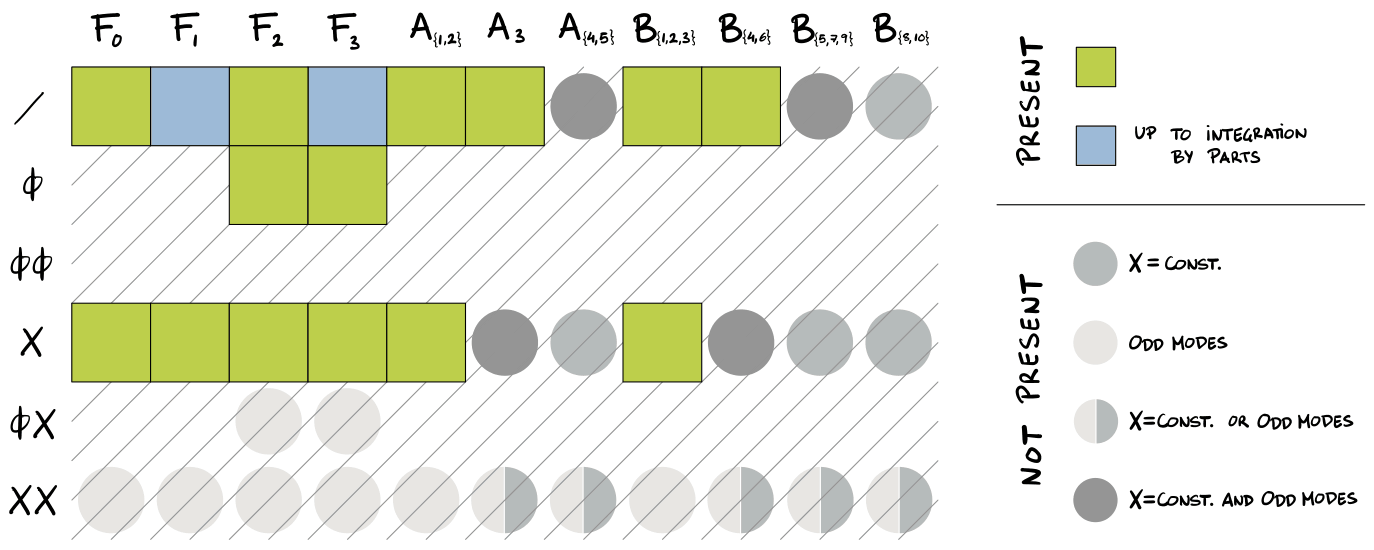}\\[0.8cm]
    \caption{Coefficients in the quadratic Lagrangian in covariant form~\eqref{eq:quadaction1}. 
    In green we show the terms that provide non-zero contributions to the covariant quadratic Lagrangian for odd-parity perturbations, Eq.~(\ref{eq:quadaction1}).
    In blue we show terms that contribute to the quadratic action only if the corresponding functions are not constants, i.e.~$F_1 , F_3 \neq \rm{const}$.
    Otherwise, they vanish up to total derivatives. The remaining coefficients are not present due to the reasons shown in the legend.
    }
    \label{fig:cubic-HOST-functions-quad-L}
\end{figure}

\begin{figure}
    \centering
    \centering
    \includegraphics[width = 0.99\textwidth]{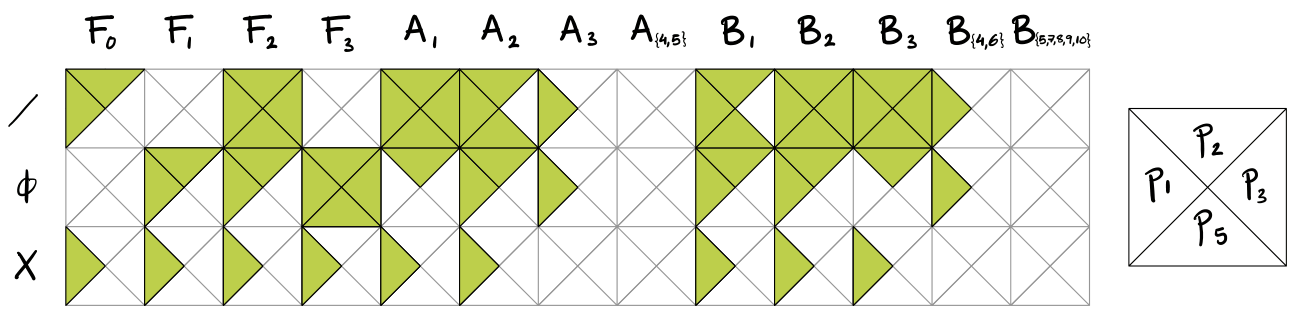}\\[0.8cm]
    \caption{Contribution from all cubic HOST functions to the different $p$-coefficients. We show in green when a specific HOST function appears in a given $p_i$ and we leave in blank the cases where they do not, e.g.~$F_0$ appears in the full expression for $p_1$ and $p_2$, but not in the one for $p_3$ and $p_4$. The full expressions can be found in the companion repository in \cite{ringdown-calculations}.
    }
    \label{fig:cubic-HOST-p-coefs}
\end{figure}

The quadratic Lagrangian for odd-parity perturbations~\eqref{eq:quadaction1} can be written in component form by substituting in the expressions for the background metric~\eqref{eq:bg_metric_Le}, the background scalar, and metric perturbations~\eqref{eq:hodd}. 
Written explicitly in terms of $h_0$ and $h_1$, we have
\begin{align}\label{L2_odd_Ein}
	\frac{2\ell + 1}{2 \pi j^2}\mathcal{L}_2 = p_1 h_0^2 + p_2 h_1^2 + p_3[(\dot{h}_1 - \partial_\rho h_0)^2+2p_4h_1\partial_\rho h_0] +p_5h_0h_1 \;,
\end{align}
where $j^2 \equiv \ell (\ell + 1)$.
Note that the background equations of motion have not been used at this stage, and we have adopted the notation in \cite{Mukohyama:2022skk}.
Analytic expressions for the $p$-coefficients in terms of cubic HOST functions are very extensive and are therefore not included here. They can be found and used, however, in the corresponding {\texttt {Mathematica}} notebook in \cite{ringdown-calculations}. 
Figure~\ref{fig:cubic-HOST-p-coefs} summarises the cubic HOST functions (in green) that contribute to the $p_i$ coefficients.
The fact that the green coefficients appear differently in Figures~\ref{fig:cubic-HOST-functions-quad-L} and \ref{fig:cubic-HOST-p-coefs} is due to the use of integration by parts to write the quadratic Lagrangian in component form as in Eq.~\eqref{L2_odd_Ein}, which necessarily rearranges the function content.

It is interesting to highlight that in our general setup without assuming shift and reflection symmetries and even without using the background equations of motion we find that
\begin{equation}
    p_4=0\;.
\end{equation}
In other words, for a static and spherically symmetric metric background (not necessarily stealth) with a linearly time-dependent scalar field, the specific combination of $h_0(\tau,\rho)$ and $h_1(\tau,\rho)$ associated with $p_4$ does not appear in cubic HOST theories even without imposing the degeneracy conditions.
This happens presumably due to the specific structure of the cubic HOST Lagrangian.
Indeed, the $p_4$ term is present in general in the context of EFT of BH perturbations with a timelike scalar profile~\cite{Mukohyama:2022skk}, which encompasses HOST theories in principle.
It was also shown in \cite{Mukohyama:2022skk} that the presence of $p_4$ forbids the existence of slowly rotating BH solutions (or otherwise leads to a diverging sound speed).
In this sense, the vanishing of $p_4$ in cubic HOST is phenomenologically desirable.

In the following Subsections, we will explicitly show expressions for the $p_i$ coefficients under several existence conditions, and identify the presence of any potential deviations from GR.
We will show here the expressions for the $p_i$ coefficients for general cubic theories (i.e.~$\textbf{Cubic}_{\text{GR-mat/GR-vac/SdS/Schw}}$), and an exhaustive list for shift-symmetric and/or quadratic theories can be found in Appendix~\ref{app-pcoeffs}.
Moreover, the beyond-GR parameter(s) for all models will be summarised in Table~\ref{beyond-GR-params-table}.

\subsection{General stealth GR with minimally coupled matter}
Let us first consider the case where all GR solutions in the presence of matter are required to exist.
Upon employing the conditions~\eqref{eq:existcondGRT} and specifying a SdS background, the coefficients $p_i$'s are given by
\begin{align}
    p_1&=r^2\sqrt{1-A}(j^2-2)\Mpl^2\;, & p_2&=-\frac{r^2}{\sqrt{1-A}}(j^2-2)\Mpl^2\;, & p_3&=\frac{\Mpl^2r^4}{\sqrt{1-A}}\;, & p_5=0\;.
    \label{eq:p-coefsGRmat}
\end{align}
These expressions for the $p_i$ coefficients are precisely the ones obtained in GR.
Therefore, in this case the dynamics of the odd-parity perturbations are the same as in GR, implying that ringdown observables associated with such perturbations, like the quasinormal mode frequencies, will be indistinguishable between the two cases.

\subsection{General stealth GR in vacuum}
Upon employing the conditions for the existence of stealth solutions~\eqref{eq:existcondGRV} and specifying an SdS background, the $p$-coefficients for the case~$\textbf{Cubic}_{\text{GR-vac}}$ are simplified to
\begin{equation}\label{eq:p-coefsGRvac}
    \begin{aligned}
    p_1&=r^2\sqrt{1-A}(j^2-2)(F_2-X_0F_{3\phi})\;, & p_2&=-\frac{r^2}{\sqrt{1-A}}(j^2-2)(F_2-X_0F_{3\phi})\;,\\
    p_3&=\frac{r^4}{\sqrt{1-A}}(F_2-X_0F_{3\phi})\;, & p_5&=0\;.
\end{aligned}
\end{equation}
From Eq.~\eqref{eq:p-coefsGRvac} we see that the presence of $F_{3\phi}$ shifts the value of $F_2$ in the same way for all non-zero coefficients~$p_1$, $p_2$ and $p_3$. In this sense, we here conclude that there is one independent combination of functions for the case~$\textbf{Cubic}_{\text{GR-vac}}$ corresponding to $F_2-X_0 F_{3\phi}$ controlling the behaviour of odd-parity perturbations. We can now distinguish two cases. First, for non-shift-symmetric models, these functions (and hence also their one independent combination) can be non-trivial functions of spacetime and, as a result, deviations from GR can appear. Because of this, we use the symbol $\checkmark_1$ in Table~\ref{main-table} to denote that there is one independent combination of HOST functions which governs the dynamics of odd modes in a way that might differ from GR.
As will be discussed in Section~\ref{sec:genRWeq}, this actually complicates the definition of quasinormal mode frequencies.
Second, if $F_2-X_0F_{3\phi}$ is a constant when evaluated on the background (note that this does not entail shift-symmetry, as it is possible that both $F_2$ and $F_{3\phi}$ are non-trivial functions of spacetime but $F_2-X_0F_{3\phi}$ is a constant), then all ringdown phenomenology in the odd sector will be indistinguishable from GR, albeit from a constant shift in the effective Planck mass [c.f.~\eqref{eq:p-coefsGRvac} with \eqref{eq:p-coefsGRmat}]. 
In this case, for instance, quasinormal mode frequencies will have the same numerical values as in standard GR.
In the shift-symmetric and/or quadratic limit, i.e.~for the cases~$\prescript{\text{SS}}{}{\textbf{Cubic}}_{\text{GR-vac}}$, $\prescript{\text{SS}}{}{\textbf{Quadratic}}_{\text{GR-vac}}$, and $\textbf{Quadratic}_{\text{GR-vac}}$, the contribution from the coefficient~$F_{3\phi}$ disappears and $F_2$ is the only contributing function [see \eqref{eq:p-coefsGroup1}]. In these cases, if $F_2$ is shift-symmetric we also recover GR predictions (therefore represented with a $\cross$ in Table~\ref{main-table}). However, for the case~$\textbf{Quadratic}_{\text{GR-vac}}$, $F_2$ can generally be a non-trivial function of spacetime, hence potentially sourcing deviations from GR. Because of this, this case is represented with the symbol~$\checkmark_1$ in Table~\ref{main-table}.

\subsection{Schwarzschild-de Sitter}
As we have seen, requiring the existence of SdS solutions as opposed to requiring general stealth GR solutions leads to a broader allowed region of theory space. Imposing the existence conditions~\eqref{eq:existcondSdS} for SdS spacetimes on cubic theories leads to the following $p_i$ coefficients:
\begin{equation}
    \begin{aligned}
    p_1&=r^2\sqrt{1-A}(j^2-2)\Big[F_2+X_0(F_{3\phi}+2A_1)\Big]\;, & p_2&=-\frac{r^2}{\sqrt{1-A}}(j^2-2)(F_2-X_0F_{3\phi})\;,\\
    p_3&=\frac{r^4}{\sqrt{1-A}}\Big[F_2+X_0(F_{3\phi}+2A_1)\Big]\;, & p_5&=0\;.
    \label{eq:p-coefsSdS}
\end{aligned}
\end{equation}
Inspecting Eq.~\eqref{eq:p-coefsSdS} we see that 2 independent combinations of HOST functions fully characterise the all $p_i$ coefficients, corresponding to
\begin{align}
    &F_2-X_0F_{3\phi}\;, & &A_1+F_{3\phi}\;.
\end{align}
Note that $p_1$ and $p_3$ are proportional to a linear combination of these two quantities.
 
In the shift-symmetric limit, all contributions from cubic functions disappear and hence for $\prescript{\text{SS}}{}{\textbf{Cubic}}_{\text{SdS}}$ and $\prescript{\text{SS}}{}{\textbf{Quadratic}}_{\text{SdS}}$, and actually also $\textbf{Quadratic}_{\text{SdS}}$, we obtain the coefficients given by \eqref{eq:coefsGroup2}, which are equivalent to Eq.~(3.6) in \cite{Takahashi:2021bml} for shift-symmetric quadratic theories.
In those cases, only $A_1$ survives as an additional contribution on top of $F_2$. When $F_2$ and $A_1$ are shift-symmetric (and therefore time-independent when evaluated on the background), the corresponding odd-parity quasinormal mode frequencies can be obtained by simple rescaling of those in GR~\cite{Mukohyama:2023xyf,Nakashi:2023vul}.
Written explicitly, for stealth Schwarzschild solutions, the relation is given by $\omega=\omega_{\rm GR}[F_2/(F_2+2X_0A_1)]^{3/2}$.

As mentioned before, when the shift-symmetry is not imposed, we recall that HOST functions can in principle contain explicit time dependences, which, as stated before, and discussed in more detail in Section~\ref{sec:genRWeq}, makes deriving the master equation in the odd-parity sector more challenging.

\subsection{Schwarzschild}\label{sec:p-coefsSchw}
In this Subsection we express the $p_i$ coefficients after imposing the existence conditions for Schwarzschild black hole solutions [Eq.~(\ref{eq:existcondSchw})]. They are given by
\begin{equation}\label{eq:p-coefsSchw}
    \begin{aligned}
    p_1&=r\sqrt{1-A}(j^2-2)\Big\{r[F_2+X_0(F_{3\phi}+2A_1)]-81\sqrt{2X_0^3(1-A)}\,B_1\Big\}\;,\\
    p_2&=-\frac{r^2}{\sqrt{1-A}}(j^2-2)(F_2-X_0F_{3\phi})\;,\\
    p_3&=\frac{r^4}{\sqrt{1-A}}\Big[F_2+X_0(F_{3\phi}+2A_1)\Big]\;,\\
    p_5&=0\;.
\end{aligned}
\end{equation}
In this case, $\textbf{Cubic}_{\text{Schw}}$, we find the following $3$ independent combinations of HOST functions as fully characterising the $p_i$ coefficients:
\begin{align}
    &F_2-X_0F_{3\phi}\;, & &A_1+F_{3\phi}\;, & &B_1\;.
\end{align}
Interestingly, one cubic function, $B_1$, survives in the shift-symmetric limit. Hence, $\prescript{\text{SS}}{}{\textbf{Cubic}}_{\text{Schw}}$ contains $2$ potential beyond-GR parameters (note that, as explained before, in shift-symmetric models $F_2$ is a constant and therefore not regarded as a beyond-GR parameter) in the evolution of odd-parity perturbations corresponding to $A_1$ and $B_1$.
The presence of $B_1$ here leads to a non-trivial $r$-dependent radial speed for GWs, something which will be discussed in Section~\ref{sec:stab-conds} together with the corresponding stability conditions. In \cite{Tomikawa:2021pca} odd-parity perturbations for Schwarzschild black holes in shift-symmetric cubic HOST theories were studied and, in particular, the contribution from $B_1$ to the fundamental quasinormal mode was calculated (while setting $A_1=0$).
In Section~\ref{sec:stab-conds} we show how a non-zero $B_1$ affects the radial speed of GWs.

Both $\prescript{\text{(SS)}}{}{\textbf{Quadratic}}_{\text{Schw}}$ contain only $A_1$ as an additional function and do in fact also fall into the same category where $p_i$ coefficients are given by \eqref{eq:coefsGroup2} (i.e.~Eq.~(3.6) in \cite{Takahashi:2021bml}, which was found for shift-symmetric quadratic HOST). 
As such, the same discussion below Eq.~(\ref{eq:p-coefsSdS}) related to the relation between quasinormal mode frequencies in this case and the ones for Schwarzschild black holes in GR also applies here.
As mentioned in the previous cases, when HOST functions above contain an implicit time dependence, the master equation for odd-parity perturbations cannot be converted to an ODE, and as a PDE, it makes the definition of quasinormal modes ambiguous.
This will be discussed in more detail in Section~\ref{sec:genRWeq}.

To conclude this Section, we have obtained the $p_i$ coefficients in the quadratic Lagrangian~\eqref{L2_odd_Ein} for odd-parity perturbations in general cubic theories under the existence conditions for stealth solutions, i.e.~$\textbf{Cubic}_{\text{GR-mat/GR-vac/SdS/Schw}}$.
As mentioned earlier, an exhaustive list of the $p_i$ coefficients for the remaining cases can be found in Appendix~\ref{app-pcoeffs}.
The beyond-GR parameter(s) for each case are summarised in Table~\ref{beyond-GR-params-table}.

\begin{table}%[H]
  \centering
  \begin{tabular*}{\textwidth}{@{\extracolsep{\fill}} llc}
    \textbf{Model} & \textbf{Beyond-GR parameter(s)} & \textbf{Symbol} \\
    \midrule
    \makecell[l]{$\prescript{\text{(SS)}}{}{\textbf{Cubic}}_{\text{GR-mat}}$, $\prescript{\text{(SS)}}{}{\textbf{Quadratic}}_{\text{GR-mat}}$, \\ $\prescript{\text{SS}}{}{\textbf{Cubic}}_{\text{GR-vac}}$, $\prescript{\text{SS}}{}{\textbf{Quadratic}}_{\text{GR-vac}}$} 
    & None & $\cross$\\
    \midrule
    $\textbf{Cubic}_{\text{GR-vac}}$ & $F_2-X_0F_{3\phi}$ at $(\phi,X)=(\phi_0,X_0)$ & $\checkmark_1$\\
    \midrule
    $\textbf{Cubic}_{\text{SdS}}$ & $F_2-X_0F_{3\phi}$, $A_1+F_{3\phi}$ at $(\phi,X)=(\phi_0,X_0)$ & $\checkmark_2$\\
    \midrule
    $\textbf{Cubic}_{\text{Schw}}$ & $F_2-X_0F_{3\phi}$, $A_1+F_{3\phi}$, $B_1$ at $(\phi,X)=(\phi_0,X_0)$ & $\checkmark_3$\\
    \midrule
    $\textbf{Quadratic}_{\text{GR-vac}}$ & $F_2$ at $(\phi,X)=(\phi_0,X_0)$ & $\checkmark_1$\\
    \midrule
    $\textbf{Quadratic}_{\text{SdS}}$, $\textbf{Quadratic}_{\text{Schw}}$ & $F_2$, $A_1$ at $(\phi,X)=(\phi_0,X_0)$ & $\checkmark_2$\\
    \midrule
    \makecell[l]{$\prescript{\text{SS}}{}{\textbf{Cubic}}_{\text{SdS}}$, $\prescript{\text{SS}}{}{\textbf{Quadratic}}_{\text{SdS}}$, \\ $\prescript{\text{SS}}{}{\textbf{Quadratic}}_{\text{Schw}}$} 
    & $A_1$ at $X=X_0$ & $\checkmark_1$\\
    \midrule
    $\prescript{\text{SS}}{}{\textbf{Cubic}}_{\text{Schw}}$ & $A_1$, $B_1$ at $X=X_0$ & $\checkmark_2$\\
    \bottomrule
  \end{tabular*}
  \caption{
  Here we collect the number of independent combinations of beyond-GR parameters for all different models (i.e.~for each set of $\{\textit{symmetry},\textit{theory},\textit{stealth solution}\}$). The last column shows the symbol we use to condense this information in Table \ref{main-table}. Note how the interpretation for $F_2$ differs for shift- vs.~non-shift-symmetric models, where in the former, as a constant, does not count as a potential deviation from GR while in the latter, as potentially depending non-trivially on spacetime coordinates, is included as a beyond-GR parameter.}
  \label{beyond-GR-params-table}
\end{table}

\section{Stability and speeds of odd-parity perturbations}\label{sec:stab-conds}
In the previous sections we have derived the conditions under which different stealth black hole solutions with time-dependent scalar hair exist for cubic HOST theories (Section~\ref{sec:background}) and obtained the quadratic Lagrangian for odd-parity perturbations (Section~\ref{sec:perturbations}). These kinds of solutions are known to exist for several large classes of ST theories (see e.g.~\cite{Babichev:2012re,Babichev:2013cya,Kobayashi:2014eva,Babichev:2016fbg,Babichev:2016kdt,Babichev:2017guv,Minamitsuji:2018vuw,BenAchour:2018dap,Motohashi:2019sen,Charmousis:2019vnf,Takahashi:2020hso,Bakopoulos:2023fmv}).
Note in passing that stealth solutions in DHOST theories have been shown to generically suffer from instability or strong coupling issues when the even sector is taken into account~\cite{deRham:2019gha,Khoury:2020aya,Takahashi:2021bml} (see Table~I in \cite{Sirera:2024ghv} for a more comprehensive summary). In this section, \blue{as a first step towards evaluating the overall stability of the different models considered,} we derive the conditions that HOST functions need to satisfy in order for linear odd-perturbations to remain stable. \blue{In the EFT context, as an initial step towards deriving generalised master equations for the even sector without strong coupling issues, the dynamics of monopole ($\ell=0$) even perturbations have been studied with the inclusion of the scordatura term, whose presence allows for the avoidance of strong coupling issues~\cite{Mukohyama:2025owu}.}

\subsection{Stability conditions}
In order to rewrite the quadratic Lagrangian~\eqref{L2_odd_Ein} in terms of one variable, 
we introduce an auxiliary field~$\chi$ and integrate out the variables~$h_0$ and $h_1$.
For a detailed description of the procedure, see, e.g.~\cite{Takahashi:2021bml,Mukohyama:2022skk}.
We have seen in the previous section that $p_4=p_5=0$ in cubic HOST under the existence conditions for stealth solutions, and therefore the corresponding terms in Eq.~\eqref{L2_odd_Ein} will be ignored in what follows.
After some manipulations, we obtain the quadratic Lagrangian for $\chi$ as\footnote{\blue{The field~$\chi$ is an effective combination of $h_0$ and $h_1$, capturing the one-degree-of-freedom nature of perturbations in the odd-parity sector. More concretely, for $p_4=0$, we have $\chi=\dot{h}_1-\partial_\rho h_0$.}}
\begin{align}\label{eq:L2-scoefs}
\frac{(j^2 - 2)(2\ell +1)}{2\pi j^2} \mathcal{L}_2 = s_1 \dot{\chi}^2 - s_2 (\partial_\rho \chi)^2 - s_3 \chi^2 \;,
\end{align}
where the parameters~$s_i$'s are defined as 
\begin{align}\label{eq:s-coefs}
    s_1 \equiv -\frac{(j^2 -2)p_3^2}{p_2} \;, \qquad s_2 \equiv \frac{(j^2 - 2)p_3^2}{p_1} \;, \qquad s_3 \equiv (j^2 - 2)p_3 \bigg[1 - \bigg(\frac{\dot{p}_3}{p_2}\bigg)^{\boldsymbol{\cdot}} - \partial_\rho \bigg(\frac{\partial_\rho p_3}{p_1}\bigg)\bigg] \;.
\end{align}
Note that these expressions apply for the case when the $p_i$-coefficients are generic functions of $\tau$ and $\rho$, and hence generalise the expressions in \cite{Khoury:2020aya,Takahashi:2021bml,Mukohyama:2022skk}
which were derived under the condition that $p_i$'s are functions only of $r=r(\rho-\tau)$.

The variable~$\chi$ now represents the propagating degree of freedom in the odd-parity sector. As usual, we define the sound speed squared of GWs (i.e.~the speed of GWs) in the radial and angular directions as\footnote{Precisely speaking, the quantities~$c_\rho^2$ and $c_\theta^2$ correspond to the sound speed of GWs in unit of the speed of photons which are assumed to be minimally coupled to gravity.}
\begin{align}
    c_\rho^2 = \frac{\bar{g}_{\rho\rho}}{|\bar{g}_{\tau\tau}|} \frac{s_2}{s_1} \;, \qquad c_\theta^2 = \lim_{\ell \rightarrow \infty} \frac{r^2}{|\bar{g}_{\tau\tau}|} \frac{s_3}{j^2 s_1}   \;. \label{eq:c_theta_M6}
\end{align}
The absence of ghost and gradient instabilities requires that $s_1$, $c_\rho^2$, and $c_\theta^2$ are positive definite:
\begin{align}
    s_1 >0\;, \qquad c_\rho^2>0\;, \qquad c_\theta^2 >0 \;.
\end{align}
Using the following parametrisation to characterise deviations from unity (i.e.~the GR prediction) in the propagation speeds,
\begin{align}
    c_\rho^2 = 1+\alpha_T^{(\rho)}\;, \qquad c_\theta^2 = 1+\alpha^{(\theta)}_{T}\;,
    \label{eq:alpha_t-def}
\end{align}
the stability requirements~$c_{\rho/\theta}^2>0$ are equivalent to $\alpha_{T}^{(\rho/\theta)}>-1$ and the GR result is obtained when $\alpha_{T}^{(\rho/\theta)}=0$.

We can now assess what these stability conditions imply for all the different cases in Table~\ref{main-table}.
As we have seen, the cases~$\prescript{\text{(SS)}}{}{\textbf{Cubic}}_{\text{GR-mat}}$ and $\prescript{\text{(SS)}}{}{\textbf{Quadratic}}_{\text{GR-mat}}$ recover GR results at the level of linear odd-parity perturbations and hence stability conditions are automatically satisfied.

For general GR solutions in vacuum, more concretely for $\textbf{Cubic}_{\text{GR-vac}}$, we obtain the following stability criterion:
\begin{equation}
    F_2-X_0F_{3\phi}>0\;.
    \label{eq:stab-conds-GR-vac}
\end{equation}
From the above we see that in the cases~$\prescript{\text{SS}}{}{\textbf{Cubic}}_{\text{GR-vac}}$ and $\prescript{\text{(SS)}}{}{\textbf{Quadratic}}_{\text{GR-vac}}$ the stability condition becomes  
\begin{equation}
    F_2>0\;.
    \label{eq:stab-conds-GR-vac-2}
\end{equation}

For the case~$\textbf{Cubic}_{\text{SdS}}$, the stability of perturbations requires that
\begin{align}
    F_2-X_0F_{3\phi}>0\;, \qquad F_2+X_0(F_{3\phi}+2A_1)>0 \;.
    \label{eq:stab-conds-group3}
\end{align}
Similarly, the stability conditions for cases~$\prescript{\text{SS}}{}{\textbf{Cubic}}_{\text{SdS}}$ and $\prescript{\text{(SS)}}{}{\textbf{Quadratic}}_{\text{SdS}}$ can be straightforwardly obtained from the condition above. In fact, the same conditions also apply for cases $\prescript{\text{(SS)}}{}{\textbf{Quadratic}}_{\text{Schw}}$, and these are given by
\begin{align}
    F_2>0\;, \qquad F_2+2X_0A_1>0 \;.
    \label{eq:stab-conds-group4}
\end{align}

This leaves us with two remaining cases.
First, we have $\textbf{Cubic}_{\text{Schw}}$, from which we obtain the following stability conditions:
\begin{align}
    \begin{split}
    &F_2-X_0F_{3\phi}>0\;, \qquad
    F_2+X_0(F_{3\phi}+2A_1)>0\;, \\
    &r[F_2+X_0(F_{3\phi}+2A_1)]-81\sqrt{2X_0^3(1-A)}\,B_1>0\;,
    \end{split}\label{eq:stab-conds-Schw}
\end{align}
where the last condition was derived from requiring $c_\rho^2>0$.
Second, the stability conditions for $\prescript{\text{SS}}{}{\textbf{Cubic}}_{\text{Schw}}$ can be straightforwardly obtained by imposing shift symmetry to the condition~(\ref{eq:stab-conds-Schw}),
\begin{align}
    F_2>0 \;, \qquad F_2+2X_0A_1>0 \;, \qquad r(F_2+2X_0A_1)-81\sqrt{2X_0^3(1-A)}\,B_1>0 \;.
    \label{eq:stab-conds-group5}
\end{align}
From the above expressions, one sees that setting the cubic HOST function to zero one recovers the condition~\eqref{eq:stab-conds-group4} for cases~$\prescript{\text{(SS)}}{}{\textbf{Quadratic}}_{\text{Schw}}$.

\subsection{Speed of gravity}
In this Subsection, for illustrative purposes, we analyse the speed of GWs in the odd-parity sector.
Let us focus on the cases~$\prescript{\text{(SS)}}{}{\textbf{Cubic}}_{\text{Schw}}$ since they contain non-trivial deviations from GR in the $p_i$ coefficients. 
Using Eqs.~(\ref{eq:p-coefsSchw}) and (\ref{eq:s-coefs}) in Eq.~(\ref{eq:c_theta_M6}), the parameters~$\alpha_T^{(\rho/\theta)}$ defined in Eq.~(\ref{eq:alpha_t-def}) are given by
\begin{align}
    \alpha_T^{(\rho)}=\frac{-2rX_0(A_1+F_{3\phi})+81\sqrt{2X_0^3(1-A)}\,B_1}{r[F_2+X_0(2A_1+F_{3\phi})]-81\sqrt{2X_0^3(1-A)}\,B_1}\;, \qquad
    \alpha_{T}^{(\theta)}=-\frac{2X_0(A_1+F_{3\phi})}{F_2+X_0(2A_1+F_{3\phi})}\;.
    \label{eq:alpha_T-Schw}
\end{align}
From the expressions for the $\alpha_T^{(\rho/\theta)}$ parameters above, it is interesting to understand how they are related to current observational constraints, most notably from the %GW170817 neutron star merger
GW event~GW170817 and the gamma-ray burst~170817A emitted from a binary neutron star merger, which constrained $|c_{\rm GW}-1|\lesssim 10^{-15}$ at the frequency scales probed by LIGO-Virgo-KAGRA (LVK)~\cite{2041-8205-848-2-L14,2041-8205-848-2-L15,TheLIGOScientific:2017qsa,LIGOScientific:2017ync,LIGOScientific:2017zic}.

The implication of the bound on GW speed on HOST theories in the context of cosmology is as follows.
On a homogeneous and isotropic cosmological background described by the Friedmann-Lema{\^i}tre-Robertson-Walker metric, the deviation of $c_{\rm GW}^2$ from unity is given by~\cite{Langlois:2017mxy}
    \begin{align}
    \alpha_T^{\text{cosm}}\equiv c_{\rm GW}^2-1
    =-\frac{2X[A_1+F_{3\phi}-6HX(B_2+B_3)]}{F_2+X[2A_1+F_{3\phi}-12HX(B_2+B_3)]}\;, \label{alphaT_cosm}
    \end{align}
where $H$ denotes the Hubble parameter.
In order for $\alpha_T^{\text{cosm}}$ to vanish irrespective of the matter content of the Universe (i.e.~irrespective of how $H$ evolves), one requires that
    \begin{align}
    A_1+F_{3\phi}=0\;, \qquad
    B_2+B_3=0 \;. \label{LV_constraint_FLRW}
    \end{align}
Having said that, note that non-trivial $\alpha_T^{\text{cosm}}$ is not ruled out for theories whose cutoff scale is lower than (or close to) the frequencies probed by LVK observations.
More specifically, higher-order operators suppressed by the scale~$\Lambda_3 \equiv (\Mpl H_0^2)^{1/3}$ (chosen so that these operators give ${\cal O}(1)$ contributions to cosmological dynamics), with $H_0$ being the present Hubble parameter, lead to a cutoff close to or below the LVK frequency band (see \cite{deRham:2018red} for theoretical background and \cite{Harry:2022zey,LISACosmologyWorkingGroup:2022wjo,Baker:2022eiz,Sirera:2023pbs,Atkins:2024nvl} for related GW phenomenology and constraints).

\begin{figure*}
    \centering
    \includegraphics[width = 0.99\textwidth]{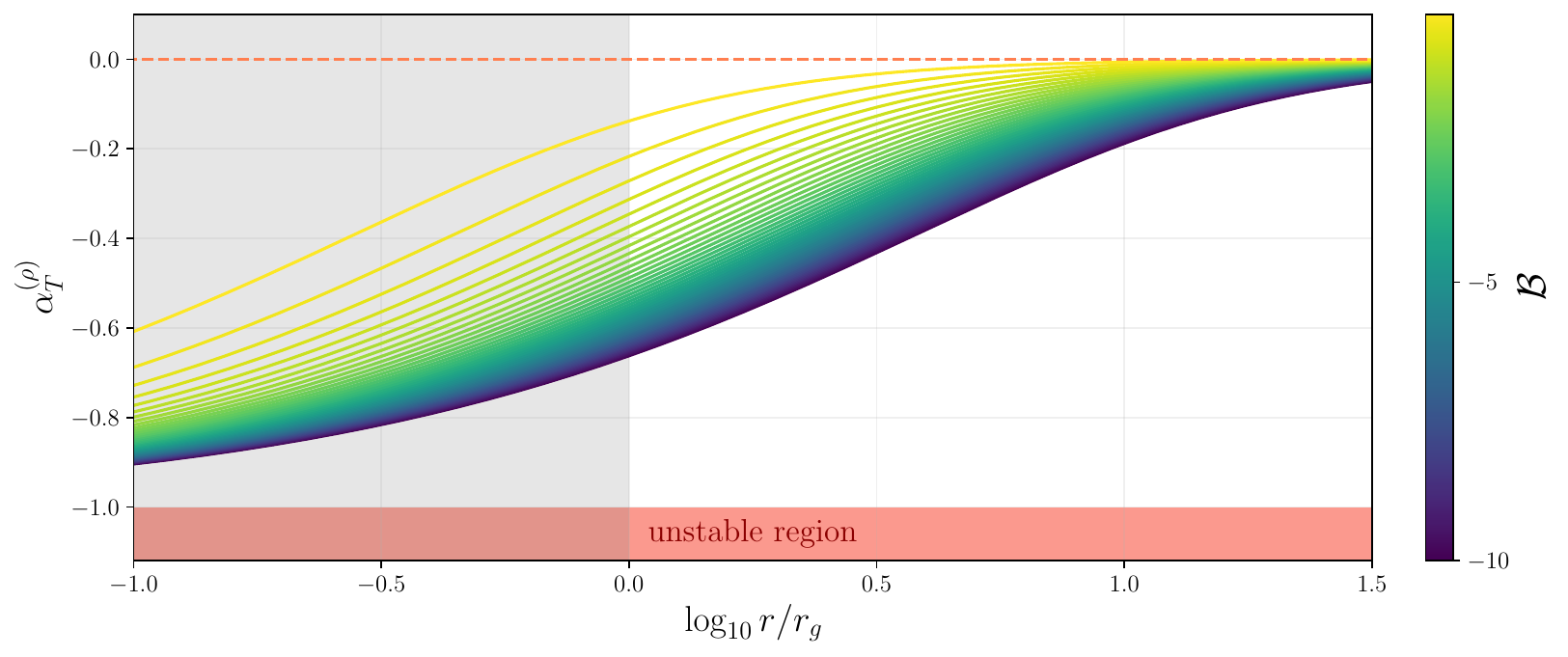}
    \caption{Speed of GWs in the radial direction as a function of distance for different values of the beyond-GR parameter~$\mathcal{B}$~\eqref{eq:alpha_T}. The distance is plotted relative to the graviton horizon radius~$r_g$, with the region inside $r_g$ shaded in grey (note that $r_g>r_s$ for ${\cal B}<0$).
    Shaded in red is the region where stability conditions are violated. In dashed orange we show the GR prediction (i.e.~$\alpha_T^{(\rho)}=0$).
    }
    \label{fig:alpha_T}
\end{figure*}

We can in principle also apply the bound on GW speed in our context, provided that the spacetime is described by the stealth Schwarzschild solution throughout the propagation of GWs.
As previously discussed in, e.g.~\cite{Mukohyama:2023xyf,Mukohyama:2024pqe}, while applying the GW170817 constraint suggests that $\alpha^{(\rho)}_T$ must approach zero at large distances, it may nonetheless have non-trivial configurations at short distances (i.e.~in the black hole environment).
In order to provide tangible results, we now assume that the relevant HOST functions are constant when evaluated at the background, i.e.~$F_2$, $F_{3\phi}$, $A_1$, $B_1$ are all constant.\footnote{Recall that we consider here non-shift-symmetric theories in general, and therefore time dependence can show up through the explicit $\phi$-dependence.
However, even if the HOST functions are not strictly constant, assuming that the scalar field is responsible for dark energy for instance, one would expect their timescale of evolution to be much longer than the timescale associated with ringdown observables here, making the HOST functions effectively constant in such an environment.}
Then, requiring $\alpha^{(\rho)}_T$ to vanish at large $r$ imposes $A_1+F_{3\phi}=0$.\footnote{Note that this coincides with the first condition in Eq.~\eqref{LV_constraint_FLRW}.
If we adopt the second condition~$B_2+B_3=0$ in our context, this can be understood as $B_1=0$ as the existence condition~\eqref{eq:existcondSchw} for $\prescript{(\text{SS})}{}{\textbf{Cubic}}_{\text{Schw}}$ imposes $18B_1=2B_2=-B_3$.}
In that case, we can rewrite the $\alpha^{(\rho)}_{T}$ parameter in \eqref{eq:alpha_T-Schw} in the following form:
\begin{align}
    \alpha_T^{(\rho)}=\frac{\left(\frac{r_s}{r}\right)^{3/2}\mathcal{B}}{1-\left(\frac{r_s}{r}\right)^{3/2}\mathcal{B}}\;, \qquad 
    % \mathcal{B}\equiv \frac{81\sqrt{2X_0^3}\,B_1}{r_s[F_2+X_0(2A_1+F_{3\phi})]}\;,
    \mathcal{B}\equiv \frac{81\sqrt{2X_0^3}\,B_1}{r_s(F_2-X_0F_{3\phi})}\;,
    \label{eq:alpha_T}
\end{align}
where, in the absence of $F_{3\phi}$, the dimensionless quantity~${\cal B}$ exactly matches that introduced in Eq.~(59) of \cite{Tomikawa:2021pca}, from which we adopt the nomenclature.
We plot $\alpha_T^{(\rho)}$ as a function of $r/r_g$ for some (negative) values of ${\cal B}$ in Figure~\ref{fig:alpha_T}, where $r_g$ corresponds to the graviton horizon (i.e.~the horizon for the odd modes).
Note that, for $\alpha_T^{(\rho)}\neq0$, $r_g$ is different than the radius of the photon horizon~$r_s$ (i.e.~the horizon for particles travelling at the speed of light).\footnote{As discussed in \cite{Tomikawa:2021pca}, for ${\cal B}<0$, Eq.~\eqref{def_rg} has only one positive solution for $r_g$, which satisfies $r_g>r_s$.
For $0<{\cal B}<{\cal B}_{\rm c}$ with ${\cal B}_{\rm c}\equiv \sqrt{108/3125}$, there exist two positive solutions to \eqref{def_rg}, of which the larger one is identified as $r_g$ and in this case we have $r_g<r_s$.
Finally, for ${\cal B}>{\cal B}_{\rm c}$, the solution for \eqref{def_rg} ceases to exist.}
Following the same discussion as in \cite{Mukohyama:2023xyf,Mukohyama:2024pqe}, one can show that $r_g$ satisfies
    \begin{align}
    A(r_g)+\alpha_T^{(\rho)}(r_g)=0\;.
    \end{align}
Specifically, on a Schwarzschild background where $A(r)=1-r_s/r$, the two horizons are related via
\begin{equation}
    r_s=[1+\alpha_T^{(\rho)}(r_g)]r_g \;.
    \label{def_rg}
\end{equation}
In Figure~\ref{fig:alpha_T} we see that $\alpha_T^{(\rho)}$ transitions smoothly from $-1$ as $r\rightarrow0$ and to zero as $r\rightarrow\infty$. 
In other words, as we approach the singularity ($r \to 0$) the odd modes `freeze', i.e.~$c_\rho=0$, while at spatial infinity we recover the GR prediction, i.e.~$c_\rho=c_{\rm light}\,(\equiv 1)$.
As a first approximation, assuming the constraint on GW speed from the event~GW170817 applies to black holes, and taking them to imply that $|\alpha_T^{(\rho)}|\lesssim 10^{-15}$ at $r\approx 10^{20}r_s$,\footnote{The event~GW170817 was observed at a distance of $40^{+8}_{-14}$~Mpc and the total mass of the binary neutron stars was approximately $2.8M_\odot$~\cite{LIGOScientific:2017zic}, with $M_\odot$ being the mass of the Sun, i.e.~$1.99\times 10^{30}$~kg.
The distance in units of $r_s$ (evaluated with the total mass of the system) is given by $r\approx 1.5\times 10^{20}r_s$.}
then we can conclude that roughly $|\mathcal{B}|\lesssim 10^{15}$, suggesting that such a measurement is not really effective for constraining this theoretical setup.

\section{Master equation for odd-parity perturbations}\label{sec:genRWeq}
In order to further connect the models investigated here with actual ringdown GW observables (e.g.~quasinormal mode frequencies), one first needs to derive a master equation for odd modes from the quadratic Lagrangian~\eqref{eq:L2-scoefs}. 
In this Section, we discuss how this derivation becomes more subtle once HOST functions are allowed to carry explicit time dependencies, as is the case for non-shift-symmetric models.

In order to clarify the subtlety, let us follow the derivation of such a master equation in e.g.~\cite{Takahashi:2021bml,Mukohyama:2022skk,Mukohyama:2023xyf} until some difficulty arises.
First, let us rewrite the quadratic Lagrangian~\eqref{eq:L2-scoefs} in terms of the Schwarzschild coordinates~$\{t, r, \theta, \phi\}$.
After some manipulations, we obtain
\begin{align}\label{eq:L2_odd_a}
	\frac{(j^2 - 2)(2\ell + 1)}{2 \pi j^2} %\mathcal{L}_2 
    \tilde{\mathcal{L}}_2 = a_1 (\partial_t \chi)^2 - a_2 (\partial_r \chi)^2 + 2a_3 (\partial_t \chi) (\partial_r \chi) - a_4\chi^2 \;,
\end{align}
with
\begin{equation}\label{eq:a-coefs}
    \begin{split}
	&a_1=\frac{s_1-(1-A)^2s_2}{\sqrt{A^3B(1-A)}}\;, \qquad
	a_2=\sqrt{\frac{B(1-A)}{A}}(s_2-s_1)\;, \\
	&a_3=\frac{(1-A)s_2-s_1}{A}\;, \qquad
	a_4=\sqrt{\frac{A}{B(1-A)}}s_3\;.
	\end{split}
\end{equation}
As mentioned earlier, though we mainly focus on the S(dS) background in the present paper, here we have kept the metric functions~$A$ and $B$ independent for generality.
Note that we have taken into account the Jacobian determinant associated with the coordinate transformation: $\tilde{\cal L}_2\equiv |\partial(\tau,\rho)/\partial(t,r)|{\cal L}_2$.

Notice that the Lagrangian~\eqref{eq:L2_odd_a} contains the cross term~$(\partial_t \chi) (\partial_r \chi)$. In the case where all the $a_i$ coefficients defined above are static with respect to the Killing time~$t$ for the background metric, one can remove the cross term by performing a transformation of the time coordinate, and it is straightforward to define a tortoise coordinate associated with the effective metric for odd modes. Note that this is the case for all shift-symmetric theories, but can also be applied in non-shift-symmetric cases as long as all HOST functions appearing in these coefficients do not carry an implicit $t$-dependence. In this case, the master equation in the form of a wave equation (i.e., the generalised Regge-Wheeler equation) was obtained in \cite{Mukohyama:2022skk,Mukohyama:2023xyf} in the context of EFT of BH perturbations.

However, the situation changes for non-shift-symmetric models, where the $a_i$ coefficients in \eqref{eq:a-coefs} are generic functions of $\tau$ and $\rho$.
In these cases, these coefficients in Schwarzschild coordinates are no longer independent of $t$, and therefore a redefinition of $t$ that removes the cross term does not exist in general.
Moreover, the position of the graviton horizon would also be $t$-dependent, which would make the definition of the tortoise coordinate subtle.
As a result, it is not possible to write the master equation in the standard form of the generalised Regge-Wheeler equation as in \cite{Mukohyama:2022skk,Mukohyama:2023xyf}. In other words, the master equation cannot be converted from a PDE to a simple ODE for a mode with a fixed frequency when the coefficients are allowed to depend on $t$. Consequently, we lose a clear definition for quasinormal mode frequencies. This implies that in order to solve the PDE one would need to resort to a numerical approach. Let us also point out, nonetheless, that when the timescale of the change of the coefficients is long enough compared to that of perturbations, those coefficients in Eq.~(\ref{eq:L2_odd_a}) can be effectively regarded as constant and the effect of the time dependence may be treated perturbatively.
This is a reasonable assumption as the time dependence of the coefficients appears only through the explicit dependence on $\phi$, which evolves in a cosmological time scale if, e.g., the scalar field is responsible for dark energy.
In this case, we expect that the effect of the time dependence can be treated in a perturbative manner.
A more in-depth exploration in this direction is left for future work.

%%%%%%%%%%%%%%%%%%%%%%%%%%%%%%%%%%%%%%%%%%%%%%%%%%%%%%%%%%%%%%%%%%%%%%%%
%\newpage
\section{Conclusions} \label{sec:conclusions}
In this paper, we have explored the landscape of stealth black hole solutions---i.e., solutions that remain identical to those in General Relativity (GR) despite the presence of a non-trivial scalar field---in the framework of general quadratic/cubic higher-order scalar-tensor (HOST) theories. We have considered configurations where the scalar field exhibits time-dependent hair while maintaining a constant background kinetic term. Our analysis has focused on deriving the precise conditions required for either all, or some given specific (Schwarzschild and Schwarzschild-de Sitter), stealth solutions to exist. Furthermore, we have examined the behaviour of odd-parity perturbations in all different cases and derived the conditions that ensure such perturbations remain stable. Table~\ref{main-table} collects the existence conditions, the nature of odd-parity perturbations, and the stability conditions for all combinations of theory setups with different classes of stealth solutions. Our key findings are:
\begin{itemize}
    \item Requiring all GR solutions to exist in the presence of generic matter leads to odd modes that display the same behaviour as in GR. Within the context of general cubic HOST theories, in order to encounter departures from GR one therefore requires one of the following: %1)
    a)~non-stealth metric solutions (see e.g.~\cite{Babichev:2013cya,Kobayashi:2014eva,Babichev:2016fbg,Babichev:2017guv}), b)~scalars with a non-constant kinetic term (see e.g.~\cite{Minamitsuji:2019tet,Bakopoulos:2023fmv,Sirera:2024ghv}), c)~study the even sector (see \cite{Takahashi:2021bml}), or d)~relax the requirement of the existence of all GR solutions in the presence of matter and employing less restrictive conditions (this leads to the conclusions in the following 2 bullet points). We leave detailed further explorations on these directions as future work.
    \item Requiring all GR solutions to exist in vacuum also leads to similar results. In the generic case of cubic HOST, we have found one potential beyond-GR parameter in the odd sector.
    However, imposing shift-symmetry and/or restricting to quadratic interactions recovers the standard GR form for the evolution of odd-parity perturbations. In these cases, departures from GR can also only appear if one takes any of the options~a)--c) spelled out in the previous bullet point.
    \item When requiring specific (Schwarzschild and Schwarzschild-de Sitter) stealth solutions to exist, we have found that a large plethora of the (theory + stealth solution) combinations considered in this paper results in a reduced set of potential deviations from GR in the odd modes, as described in Table~\ref{main-table}. In most cases, odd-parity quasinormal mode frequencies can be obtained from those in GR via a simple rescaling, in the same fashion as in \cite{Mukohyama:2023xyf,Nakashi:2023vul}.
    \item We have identified a unique deviation from GR that does not fall in the previous category for Schwarzschild black holes in cubic HOST (shift-symmetric or otherwise). We have shown how this deviation, denoted by the parameter~$\mathcal{B}$, modifies the propagation speed of odd modes in a non-trivial way, in particular with an $r$-dependent $\alpha_T$ parameter.
    We have shown that the speed of gravity is modified in the black hole environment (while still satisfying stability conditions) and approaches the speed of light at cosmological distances, hence making this an interesting healthy model in light of the constraint on the propagation speed of gravitational waves (GWs) from the event~GW170817. The fundamental quasinormal mode frequency for a shift-symmetric version of this model was investigated in \cite{Tomikawa:2021pca}. Given the uniqueness of the $\mathcal{B}$~signature in the large class of models we studied, a further investigation of the quasinormal mode spectrum with contributions from $\mathcal{B}$, as well as an assessment of the observability of $\mathcal{B}$ by current and future GW detectors constitute interesting directions for future research.
    \item We have established that $p_4=0$ in the quadratic Lagrangian~\eqref{L2_odd_Ein} for odd-parity perturbations about static and spherically symmetric background
    in general cubic HOST theories. In general, non-zero $p_4$ appears in the context of EFT of BH perturbations with a timelike scalar profile~\cite{Mukohyama:2023xyf}. We have shown here that in covariant cubic HOST theories, contributions to $p_4$ from individual terms cancel out in a non-trivial manner. The relevance of this result is that the presence of $p_4\neq0$ is associated with the exclusion of slowly rotating black hole solutions (or otherwise the divergence of the radial sound speed at spatial infinity)~\cite{Mukohyama:2023xyf}, and so we show that cubic HOST theories do not suffer from this problem.
    \item In the context of non-shift-symmetric theories, we have shown how the master equation for odd-parity perturbations cannot be written in the standard ODE form of the generalised Regge-Wheeler equation due to the time-dependence of its coefficients. We have argued that solving the corresponding PDE would require numerical methods and/or further approximations. We leave a more detailed analysis for future work.
    \item When we require the degeneracy conditions~\eqref{DC_quadratic}, \eqref{DC_cubic}, and \eqref{DC_quadratic+cubic} for quadratic/cubic DHOST theories of class~N-I~\cite{BenAchour:2016fzp} as well as the existence conditions derived in this work, we have concluded that the cubic DHOST part of the theory is not allowed.
    (The authors of \cite{Minamitsuji:2019shy} already pointed out that the existence conditions for stealth Schwarzschild(-de Sitter) solutions are not compatible with cubic DHOST.)
    In this case, we are left with the quadratic DHOST of class~${}^{2}$N-I, for which the compatibility between the degeneracy and existence conditions is guaranteed.
\end{itemize}

In the present paper we have shown how restrictive it is to require the existence of exact stealth solutions with timelike scalar profile in the space of scalar-tensor theories. On the other hand, as briefly reviewed in the introduction, one needs to introduce so-called scordatura terms in order to avoid strong coupling of perturbations, promoting the background solution to an approximately stealth one that behaves as stealth for any practical purposes at the level of the background and that is free from the strong coupling issue for perturbations. Since scordatura terms are of order unity (and not necessarily small) in the unit of the cutoff of the theory, it is expected that relaxing the existence condition of exact stealth solutions to that of approximately stealth ones should significantly broaden the space of scalar-tensor theories. We leave detailed investigations of this important problem for future work.

%%%%%%%%%%%%%%%%%%       Acknowledgements          %%%%%%%%%%%%%%%%%%%%
\newpage
\section*{Acknowledgements}
Sergi Sirera is grateful for the kind hospitality extended to him at YITP.
This work was supported by World Premier International Research Center Initiative (WPI), MEXT, Japan.
HK is supported by JST SPRING, Grant No.~JPMJSP2110.
SM is supported in part by JSPS KAKENHI Grant No.~JP24K07017.
JN is supported by an STFC Ernest Rutherford Fellowship (ST/S004572/1).
SS is supported by a JSPS Postdoctoral Fellowship PE24719 and an STFC studentship.
KT is supported in part by JSPS KAKENHI Grant No.~JP23K13101.
VY is supported by grants for development of new faculty staff, Ratchadaphiseksomphot Fund, Chulalongkorn University.
In deriving the results of this paper, we have used xAct~\cite{xAct}.

For the purpose of open access, the authors have applied a Creative Commons Attribution (CC BY) licence to any Author Accepted Manuscript version arising from this work.
\\

\noindent{\bf Data availability} The results presented here and all the information necessary to fully reproduce them can be found at~\cite{ringdown-calculations}.

%%%%%%%%%%%%%%%%%%       Appendix          %%%%%%%%%%%%%%%%%%%%

\appendix
\section{Exhaustive list of existence conditions}\label{app-existconds}
In the main text we have explicitly included the existence conditions for the following cases: $\textbf{Cubic}_{\text{GR-mat}}$~\eqref{eq:existcondGRT}, $\textbf{Cubic}_{\text{GR-vac}}$~\eqref{eq:existcondGRV}, $\textbf{Cubic}_{\text{SdS}}$~\eqref{eq:existcondSdS}, and $\textbf{Cubic}_{\text{Schw}}$~\eqref{eq:existcondSchw}. As the most general cases, these are the most interesting expressions to show, also because the conditions for simpler theories can be directly obtained from them in the appropriate limits as explained in the main text. Here, with the aim of providing a comprehensive review, we list the conditions for the remaining cases in Table~\ref{exist-conds-table}.

Some of the expressions shown in the Table have already been obtained in previous work. The conditions for $\prescript{\text{(SS)}}{}{\textbf{Quadratic}}_{\text{GR-mat}}$ and $\prescript{\text{(SS)}}{}{\textbf{Quadratic}}_{\text{GR-vac}}$ can be obtained from Eq.~(2.6) in \cite{Takahashi:2020hso}. In addition, the conditions for $\prescript{\text{SS}}{}{\textbf{Cubic}}_{\text{SdS}}$ and $\prescript{\text{SS}}{}{\textbf{Cubic}}_{\text{Schw}}$ correspond to Eq.~(21) and Eq.~(18) respectively in \cite{Minamitsuji:2019shy}. Similarly, the conditions for $\prescript{\text{SS}}{}{\textbf{Quadratic}}_{\text{SdS}}$ and $\prescript{\text{SS}}{}{\textbf{Quadratic}}_{\text{Schw}}$ correspond respectively to Eqs.~(42) and (23) in \cite{Motohashi:2019sen}.
Note that when comparing these expressions with the corresponding ones in the literature one needs to take into account different conventions for $X$, which is often defined without the factor of $-1/2$.

\newcounter{tableeqn}[section]
\renewcommand{\thetableeqn}{\Alph{section}.\arabic{tableeqn}}
\newcounter{tablesubeqn}[tableeqn]
\renewcommand{\thetablesubeqn}{\thetableeqn\alph{tablesubeqn}}
\begin{table}[H]
\centering
\stepcounter{table}
\begin{tabular}{l p{11cm} r}
\textbf{Model} & \textbf{Existence conditions} & \\
\hline
$\prescript{\text{SS}}{}{\textbf{Cubic}}_{\text{GR-mat}}$ &
\begin{tabular}{lr}
$F_0 + 2\Lambda \Mpl^2 = 0$, \quad $F_2 - \Mpl^2 = 0$,\quad $F_{0X} = F_{1X} = F_{2X} = F_{3X} = 0$ \\
$A_1 = A_2 = A_{1X} = A_{2X} = A_3 = 0$, \\
$B_1 = B_{1X} = B_2 = B_{2X} = B_3 = B_{3X} = B_4 = B_6 = 0$. 
\end{tabular}  &
\refstepcounter{tableeqn} (\thetableeqn)\label{eq:existcondsGRmatSSCub}\\
\hline
$\textbf{Quadratic}_{\text{GR-mat}}$ &
\begin{tabular}{l}
$F_0 + 2\Lambda \Mpl^2 = -2X_0(F_{1\phi} + 2F_{2\phi\phi}) = X_0(F_{0X} - 2F_{2\phi\phi})$\\
$F_2 - \Mpl^2 = 0,\quad F_{1X} = F_{2\phi} = F_{2X} = A_1 = A_2 = A_{1X} = A_{2X} = A_3 = 0$.
\end{tabular}  &
\refstepcounter{tableeqn} (\thetableeqn)\label{eq:existcondsGRmatQuad}\\
\hline
$\prescript{\text{SS}}{}{\textbf{Quadratic}}_{\text{GR-mat}}$ &
\begin{tabular}{l}
$F_0 + 2\Lambda \Mpl^2 = 0$, \quad $F_2 - \Mpl^2 = 0$, \\
$F_{0X} = F_{1X} = F_{2X} = A_1 = A_2 = A_{1X} = A_{2X} = A_3 = 0$. \\
\end{tabular}  &
\refstepcounter{tableeqn} (\thetableeqn)\label{eq:existcondsGRmatSSQuad}\\
\hline
$\prescript{\text{SS}}{}{\textbf{Cubic}}_{\text{GR-vac}}$ &
\begin{tabular}{lr}
$F_0+2\Lambda F_2=0,\quad F_{0X}=-2\Lambda(2F_{2X}+X_0A_{1X}),$\quad $F_{1X}-\Lambda F_{3X}=-4\Lambda X_0 B_4,$ \\
$A_3=A_{1X}=-A_{2X},\quad 2B_4=-2B_{1X}=B_{2X},$ \\
$A_1=A_2=B_1=B_2=B_3=B_{3X}=B_6=0$.
\end{tabular}  &
\refstepcounter{tableeqn} (\thetableeqn)\label{eq:existcondsGRvacSSCub}\\
\hline
$\textbf{Quadratic}_{\text{GR-vac}}$ &
\begin{tabular}{l}
$F_0+2\Lambda F_2=-2X_0(F_{1\phi}+2F_{2\phi\phi}),$ \\
$F_{0X}=-2[F_{1\phi}+F_{2\phi\phi}+\Lambda(2F_{2X}+X_0A_{1X})],$ \\
$3F_{2\phi}+X_0F_{1X}=2X_0^2A_{3\phi},\quad F_{2\phi}=A_1=A_2=0,$\quad $A_3=A_{1X}=-A_{2X}$.
\end{tabular}  &
\refstepcounter{tableeqn} (\thetableeqn)\label{eq:existcondsGRvacQuad}\\
\hline
$\prescript{\text{SS}}{}{\textbf{Quadratic}}_{\text{GR-vac}}$ &
\begin{tabular}{l}
$F_0+2\Lambda F_2=0,\quad F_{0X}=-2\Lambda(2F_{2X}+X_0A_{1X}),$\\
$F_{1X}=0$,\quad $A_3=A_{1X}=-A_{2X}$,\quad $A_1=A_2=0$.\\
\end{tabular}  &
\refstepcounter{tableeqn} (\thetableeqn)\label{eq:existcondsGRvacSSQuad}\\
\hline
$\prescript{\text{SS}}{}{\textbf{Cubic}}_{\text{SdS}}$ &
\begin{tabular}{lr}
$F_0+2\Lambda F_2=-4\Lambda X_0 A_{1},\quad F_{0X}+4\Lambda F_{2X}=-2\Lambda[A_1+X_0(2A_{2X}+3A_3)],$ \\
$A_{1}=-A_{2},\quad A_{1X}=-A_{2X},\quad B_1=B_2=B_3=B_{3X}=0,$ \\
$F_{1X}-\Lambda F_{3X}=-4\Lambda X_0(2B_4+B_6+B_{1X}),\quad B_4+B_6=B_{1X}+B_{2X}$.
\end{tabular}  &
\refstepcounter{tableeqn} (\thetableeqn)\label{eq:existcondsSdSSSCub}\\
\hline
$\textbf{Quadratic}_{\text{SdS}}$ &
\begin{tabular}{l}
$F_0+2\Lambda F_2=-2X_0(F_{1\phi}+2F_{2\phi\phi}+2\Lambda A_{1}), \quad F_{2\phi}=2X_0A_{2\phi},$ \\
$F_{0X}+4\Lambda F_{2X}=-2[F_{1\phi}+F_{2\phi\phi}-A_1-X_0(2A_{2X}+3A_3)],$ \\
$A_{1}=-A_{2},\quad A_{1X}=-A_{2X},\quad F_{1X}=-4A_{2\phi}+2X_0 A_{3\phi}$.
\end{tabular}  &
\refstepcounter{tableeqn} (\thetableeqn)\label{eq:existcondsSdSQuad}\\
\hline
$\prescript{\text{SS}}{}{\textbf{Quadratic}}_{\text{SdS}}$ &
\begin{tabular}{l}
$F_0+2\Lambda F_2=-4\Lambda X_0 A_{1},$\\
$F_{0X}+4\Lambda F_{2X}=-2\Lambda[A_1+X_0(2A_{2X}+3A_3)],$ \\
$A_{1}=-A_{2},\quad A_{1X}=-A_{2X},\quad F_{1X}=0$.\\
\end{tabular}  &
\refstepcounter{tableeqn} (\thetableeqn)\label{eq:existcondsSdSSSQuad}\\
\hline
$\prescript{\text{SS}}{}{\textbf{Cubic}}_{\text{Schw}}$ &
\begin{tabular}{lr}
$F_0=F_{0X}=F_{1X}=0,\quad A_{1}=-A_{2},\quad A_{1X}=-A_{2X},$ \\
$18B_1=2B_2=-B_3,\quad 3B_3=X_0[9(B_{1X}+B_{2X}-B_4-B_6)+5B_{3X}]$.\\
\end{tabular}  &
\refstepcounter{tableeqn} (\thetableeqn)\label{eq:existcondsSchwSSCub}\\
\hline
$\textbf{Quadratic}_{\text{Schw}}$ &
\begin{tabular}{l}
$F_0=-2X_0(F_{1\phi}+2F_{2\phi\phi}), \quad F_{2\phi}=%-X_0(-2A_{2\phi})
2X_0A_{2\phi},$ \\
$F_{0X}=-2(F_{1\phi}+F_{2\phi\phi}),$ \\
$A_{1}=-A_{2},\quad A_{1X}=-A_{2X},\quad F_{1X}=-4A_{2\phi}+2X_0 A_{3\phi}.$
\end{tabular}  &
\refstepcounter{tableeqn} (\thetableeqn)\label{eq:existcondsSchwQuad}\\
\hline
$\prescript{\text{SS}}{}{\textbf{Quadratic}}_{\text{Schw}}$ &
\begin{tabular}{l}
$F_0=F_{0X}=F_{1X}=0,\quad A_{1}=-A_{2},\quad A_{1X}=-A_{2X}.$\\
\end{tabular}  &
\refstepcounter{tableeqn} (\thetableeqn)\label{eq:existcondsSchwSSQuad}\\
\hline

\end{tabular}
\caption{Here we collect an exhaustive list of existence conditions for all models not shown in the main text.}
\label{exist-conds-table}
\end{table}

\section{Terms in covariant quadratic Lagrangian}\label{app-covquadL}
Writing the perturbed action to quadratic order in metric perturbations as
\begin{align}
    S_{\rm grav}^{(2)}=\frac{1}{4}\int {\rm d}^4x\sqrt{-g}\sum_a\biggl[
    &\sum_{K=0}^{3}\delta\mathcal{L}_{F_{Ka}}F_{Ka}
    +\sum_{I=1}^5\delta\mathcal{L}_{A_{Ia}}A_{Ia}+\sum_{J=1}^{10}\delta\mathcal{L}_{B_{Ja}}B_{Ja}\biggr]\;,
\end{align}
where $a=\{\emptyset,\phi,X,\phi\phi,XX,\phi X\}$ corresponding to the different $\phi$- and $X$-derivatives, the only non-zero terms for odd-parity perturbations in theories where the scalar has a constant kinetic term are given below.\footnote{Precisely speaking, the coefficients presented here are those that we see just after expanding (the gravitational part of) the Lagrangian~\eqref{eq:action1} for cubic HOST theories up to the quadratic order in odd-parity perturbations. The explicit form of the Lagrangian changes when one performs integration by parts.}
These expressions can be found and directly used in the provided {\texttt {Mathematica}} files~\cite{ringdown-calculations}.
\begin{align}
    \delta\mathcal{L}_{F_{0}}&=-h_\mu^\nu h_\nu^\mu\;,\label{eq:dLF0} \\
    \delta\mathcal{L}_{F_{0X}}&=-2\phi^\mu\phi^\nu h_\mu^\sigma h_{\nu\sigma}\;, \\
    \delta\mathcal{L}_{F_{1}}&=h_{\nu\sigma}(4\phi^\mn h_\mu^\sigma-\Box\phi h^{\nu\sigma})+2\phi^\mu[2h_\mu^\nu\nabla_\sigma h_\nu^\sigma-h^{\nu\sigma}(\nabla_\mu h_{\nu\sigma}-2\nabla_\sigma h_\mn)]\;, \\
    \delta\mathcal{L}_{F_{1X}}&=-2\phi^\mu\phi^\nu\Box\phi h_\mu^\sigma h_{\nu\sigma}\;, \\
    \delta\mathcal{L}_{F_{2}}&=4R_{\mn}h^{\mu\sigma}h^\nu_\sigma-Rh_\mu^\nu h_\nu^\mu+\nabla^\sigma h^\mn(2\nabla_\nu h_{\mu\sigma}-\nabla_\sigma h_\mn)\;, \\
    \delta\mathcal{L}_{F_{2\phi}}&=4\phi^\mu[h^{\nu\sigma}(\nabla_\sigma h_\mn-\nabla_\mu h_{\nu\sigma})+h_\mu^\nu\na_\si h_\nu^\sigma]\;, \\
    \delta\mathcal{L}_{F_{2X}}&=-2\phi^\mu[R\phi^\sigma h_\mu^\nu h_{\nu\sigma}
    +(\phi^\nu\nabla^\sigma h_{\mn}+2h_{\mn}\phi^{\nu\sigma})\nabla_\rho h_\sigma^\rho]\;,\\
    \delta\mathcal{L}_{F_{3}}&=\phi^\mu\bigl\{2G^{\nu\sigma}[h_\mu^\rho(2\nabla_\sigma h_{\nu\rho}-\nabla_\rho h_{\nu\sigma})+2h_\nu^\rho(\nabla_\sigma h_{\mu\rho}+\nabla_\rho h_{\mu\sigma}-\nabla_\mu h_{\sigma\rho})]-4\nabla^\sigma h_\mu^\nu\nabla_\rho\nabla_{(\nu}h_{\sigma)}^\rho\nonumber\\
    &\qquad\qquad+\nabla_\mu h^{\nu\sigma}(2\nabla_\rho\nabla_\sigma h_\nu^\rho-\Box h_{\nu\sigma})+2\nabla_\nu h_\mu^\nu\nabla_\rho\nabla_\sigma h^{\sigma\rho}+2\nabla^\sigma h_\mu^\nu\Box h_{\nu\sigma}\nonumber\\
    &\qquad\qquad-h^{\nu\sigma}[R(\nabla_\mu h_{\nu\sigma}-2\nabla_\sigma h_{\mu\nu})+2R_{\nu\sigma}\nabla_\rho h_\mu^\rho]\bigl\}\nonumber \\
    &\qquad+\phi^{\mn}\big[4G^{\sigma\rho}h_{\mu\sigma}h_{\nu\rho}+8G_\nu^\sigma h_\mu^\rho h_{\sigma\rho}-G_{\mn}h_\sigma^\rho h_\rho^\sigma-2R_{\sigma\rho}h_{\mn}h^{\sigma\rho}+4Rh_\mu^\sigma h_{\nu\sigma}\nonumber\\
    &\qquad\qquad+\nabla_\mu h^{\sigma\rho}(4\nabla_\rho h_{\nu\sigma}-\nabla_\nu h_{\sigma\rho})-2(\nabla^\sigma h_{\mn}\nabla_\rho h_\sigma^\rho+\nabla_\rho h_{\nu\sigma}\nabla^\rho h_\mu^\sigma-\nabla_\sigma h_{\nu\rho}\nabla^\rho h_\mu^\sigma)\bigl]\nonumber\\
    &\qquad+2\phi^{\mn\sigma}\big[h_\sigma^\rho(2\nabla_\mu h_{\nu\rho}-\nabla_\rho h_{\mn})+2h_\mu^\rho(\nabla_\nu h_{\sigma\rho}-\nabla_\sigma h_{\nu\rho}+\nabla_\rho h_{\nu\sigma})-h_{\mn}\nabla_\rho h_\sigma^\rho\bigl]\nonumber\\
    &\qquad+\Box\phi\bigg[\!\left(\frac{1}{2}\nabla_\rho h_{\nu\sigma}-\nabla_\sigma h_{\nu\rho}\right)\nabla^\rho h^{\nu\sigma}-2R_{\sigma\rho}h_\nu^\rho h^{\nu\sigma}\bigg]\nonumber\\
    &\qquad+4\phi^\mu{}_{[\mn]}h^{\sigma\rho}\nabla^\nu h_{\sigma\rho}-2\phi_\mu{}^{\mn}(h^{\sigma\rho}\nabla_\rho h_{\nu\sigma}+h_\nu^\sigma \nabla_\rho h_\sigma^\rho)\;,\\
    \delta\mathcal{L}_{F_{3\phi}}&=2\phi^\mu\bigl\{\phi^{\nu\rho}[2h_\nu^\sigma(\nabla_\sigma h_{\mu\rho}+\nabla_\rho h_{\mu\sigma}-\nabla_\mu h_{\sigma\rho})+h_\mu^\sigma(2\nabla_\nu h_{\sigma\rho}-\nabla_\sigma h_{\nu\rho})] \nonumber\\
    &\qquad+\Box\phi[h^{\sigma\rho}(\nabla_\mu h_{\sigma\rho}-\nabla_\sigma h_{\mu\rho})-h_\mu^\rho \nabla_\sigma h_\rho^\sigma]\bigl\}\;, \\
    \delta\mathcal{L}_{F_{3X}}&=\phi^\mu\bigl\{2h_{\mu\nu}[\phi^\lambda\phi^{\sigma\rho}G_{\sigma\rho}h_\lambda^\nu+\phi^{\nu\sigma}\phi^{\rho\lambda}(\nabla_\sigma h_{\rho\lambda}-2\nabla_\lambda h_{\sigma\rho})+\Box\phi\phi^{\nu\sigma}\nabla_\rho h_\sigma^\rho]\nonumber\\
    &\qquad+\phi^\nu\nabla^\sigma h_{\mu\nu}[\phi^{\rho\lambda}(\nabla_\sigma h_{\rho\lambda}-2\nabla_\lambda h_{\rho\sigma})+\Box\phi\nabla_\rho h_\sigma^\rho]\bigr\}\;,\\
    \delta\mathcal{L}_{A_{1}}&=\phi^\mu\phi^\nu[\nabla_\mu h^{\sigma\rho}(\nabla_\nu h_{\sigma\rho}-4\nabla_\rho h_{\nu\sigma})+4\nabla^\rho h_\mu^\sigma\nabla_{(\sigma}h_{\rho)\nu}] \nonumber\\
    &\qquad+\phi^{\mu\nu}[h_{\sigma\rho}(8h_\mu^\rho \phi_\nu^\sigma-h^{\sigma\rho}\phi_{\mu\nu})+4h_{\mu\sigma}h_{\nu\rho}\phi^{\sigma\rho}]\nonumber\\
    &\qquad+4\phi^\mu\phi^{\nu\sigma}[h_\mu^\rho(\nabla_\sigma h_{\nu\rho}+\nabla_\sigma h_{\nu\rho}-\nabla_\rho h_{\nu\sigma})+2h_\sigma^\rho(\nabla_\rho h_{\mu\nu}-\nabla_\mu h_{\nu\rho}+\nabla_\nu h_{\mu\rho})]\;,\\
    \delta\mathcal{L}_{A_{1X}}&=-2\phi^\mu\phi^\nu h_\mu^\sigma h_{\nu\sigma}\phi^\alpha_\beta\phi_\alpha^\beta\;,\\
    \delta\mathcal{L}_{A_{2}}&=\Box\phi\bigl\{h_{\mu\nu}[-\Box\phi h^{\mu\nu}+8h_\sigma^\nu\phi^{\mu\sigma}-4\phi^\sigma(\nabla_\sigma h^{\mu\nu}-2\nabla^\nu h_\sigma^\mu)]+8\phi^\sigma h_\sigma^\mu \nabla_\nu h_\mu^\nu\bigl\}\;,\\
    \delta\mathcal{L}_{A_{2X}}&=-2\phi^\mu\phi^\nu(\Box\phi)^2h_\mu^\sigma h_{\nu\sigma}\;, \\
    \delta\mathcal{L}_{A_{3}}&=2\phi^\mu\phi^\nu\Box\phi\bigl[\phi^\rho h_\mu^\sigma(2\nabla_\rho h_{\nu\sigma}+\nabla_\sigma h_{\nu\rho})+2h_{\mu\sigma}h_{\nu\rho}\phi^{\sigma\rho}\bigl]\;,\\
    \delta\mathcal{L}_{B_{1}}&=(\Box\phi)^2h^{\mu\nu}\bigl[-\Box\phi h_{\mu\nu}+12\nabla_\sigma(\phi_\mu h^\sigma_\nu)-6\phi^\sigma(\nabla_\sigma h_{\mu\nu}-2\nabla_\nu h_{\sigma\mu})\bigl]\;,\\
    \delta\mathcal{L}_{B_{1X}}&=-2\phi^\mu\phi^\nu(\Box\phi)^3h_\mu^\sigma h_{\nu\sigma}\;,\\
    \delta\mathcal{L}_{B_{2}}&=\Box\phi \phi^\mu\phi^\nu\bigl[\nabla_\mu h^{\sigma\rho}\nabla_\nu h_{\sigma\rho}+4\nabla^\sigma h_\mu^\rho(\nabla_{(\sigma}h_{\rho)\nu}-\nabla_\nu h_{\sigma\rho})\bigl]\nonumber\\
    &\qquad+4\Box\phi \phi^{\mu\nu}\phi^\sigma\bigl[h_\mu^\rho(4\nabla_{[\nu}h_{\sigma]\rho}+\nabla_\rho h_{\sigma\nu})+2h_\sigma^\rho \nabla_{[\mu}h_{\rho]\nu}\bigl]\nonumber\\
    &\qquad+2\phi^\alpha_\beta\phi^\beta_\alpha \phi^\sigma\bigl(2h_\sigma^\nu\nabla_\mu h_\nu^\mu+2h^{\mu\nu}\nabla_\nu h_{\sigma\mu}-h^{\mu\nu}\nabla_\sigma h_{\mu\nu}\bigl)\nonumber\\
    &\qquad+h_{\mu\nu}\bigl[-\Box\phi \phi^\alpha_\beta\phi^\beta_\alpha h^{\mu\nu}+4h_\sigma^\nu(2\Box\phi \phi_\rho^\mu \phi^{\rho\sigma}+\phi^\alpha_\beta\phi^\beta_\alpha\phi^{\sigma\mu})+4\Box\phi\phi^{\mu\sigma}\phi^{\nu\rho}h_{\sigma\rho}\bigl]\;,\\
    \delta\mathcal{L}_{B_{2X}}&=-2\phi^\mu\phi^\nu\Box\phi \phi^\alpha_\beta\phi^\beta_\alpha h_\mu^\sigma h_{\nu\sigma}\;,\\
    \delta\mathcal{L}_{B_{3}}&=3\phi^{\mu\nu}\bigg\{\frac{1}{3}\phi^\sigma_\nu h_{\rho\lambda}(12\phi^\rho_\sigma h_\mu^\lambda-\phi_{\sigma\mu}h^{\rho\lambda})+4\phi^\sigma_\nu\phi^{\rho\lambda}h_{\mu\lambda}h_{\sigma\rho}+2\phi^\sigma\phi^{\rho\lambda}h_{\mu\lambda}(\nabla_\sigma h_{\nu\rho}+\nabla_\nu h_{\sigma\rho}+\nabla_\rho h_{\sigma\nu}) \nonumber \\
    &\qquad+2\phi^\sigma\phi^\rho_\nu\bigl[h_\sigma^\lambda(\nabla_\mu h_{\rho\lambda}+\nabla_\rho h_{\mu\lambda}-\nabla_\lambda h_{\mu\rho})+2h^\lambda_\mu(-\nabla_\sigma h_{\rho\lambda}+\nabla_\rho h_{\sigma\lambda}+\nabla_\lambda h_{\sigma\rho})\bigl] \nonumber \\
    &\qquad+\phi^\sigma\phi^\rho\bigl[2\nabla_{[\nu}h_{\rho]\lambda}(\nabla_\mu h_\sigma^\lambda+2\nabla^\lambda h_{\mu\sigma})+2\nabla_\rho h_{\nu\lambda}\nabla_{[\sigma}h_{\mu]}^\lambda+\nabla_\lambda h_{\rho\nu}\nabla^\lambda h_{\sigma\mu}\bigl]\bigg\}\;,\\
    \delta\mathcal{L}_{B_{3X}}&=-2\phi^\mu\phi^\nu \phi^\alpha_\beta\phi^\beta_\gamma\phi^\gamma_\alpha h_\mu^\sigma h_{\nu\sigma}\;,\\
    \delta\mathcal{L}_{B_{4}}&=2\phi^\mu\phi^\nu(\Box\phi)^2\bigl[\phi^\rho h_\mu^\sigma(2\nabla_\rho h_{\nu\sigma}+\nabla_\sigma h_{\nu\rho})+2h_{\mu\sigma}h_{\nu\rho}
    \phi^{\sigma\rho}\bigl]\;,\\
    \delta\mathcal{L}_{B_{6}}&=2\phi^\mu\phi^\nu\phi^\alpha_\beta\phi^\beta_\alpha\bigl[\phi^\rho h_\mu^\sigma(2\nabla_\rho h_{\nu\sigma}+\nabla_\sigma h_{\nu\rho})+2h_{\mu\sigma}h_{\nu\rho}
    \phi^{\sigma\rho}\bigl]\;,\label{eq:dLB6}
\end{align}
where the indices of the perturbed metric are raised/lowered by the background metric, and $R$, $R_{\mu\nu}$, $G_{\mu\nu}$, and the covariant derivatives are evaluated on the background. Also, we have used the standard definition of symmetric and antisymmetric tensors,
\begin{align}
    A_{(\mn)}&\equiv\frac{1}{2}(A_\mn+A_{\nu\mu})\;,
    \quad\quad A_{[\mn]}\equiv\frac{1}{2}(A_\mn-A_{\nu\mu})\;.
    \label{sym-tensors}
\end{align}

All other contributions are shown to be zero if one assumes $X={\rm const.}$ [i.e.~by employing the relations in Eq.~\eqref{eq:constantX}] and/or restricts to odd-parity modes, as described in Figure~\ref{fig:cubic-HOST-functions-quad-L}. Here we include the simplifying relations valid for odd-parity modes about a spherically symmetric background with the component form of Eq.~(\ref{eq:hodd}).
These relations (as well as derivatives of some of them) have also been used to simplify the expressions in Eqs.~\eqref{eq:dLF0}--\eqref{eq:dLB6}.
    \begin{align}
    &h^\mu_\mu
    =\phi^\mu\phi^\nu h_{\mu\nu}=\phi^{\mu\nu} h_{\mu\nu}
    =\phi^\mu\nabla_\nu h^\nu_\mu
    =\phi^\mu\phi^\nu\nabla_\sigma h_\mu^\sigma\nabla_\rho h_\nu^\rho
    =\phi^\mu\phi^\nu\phi^\sigma\nabla_\sigma h_{\mu\nu}=\phi^{\mu\nu}\phi^\sigma\nabla_\nu h_{\sigma\mu}
    =0\;.
    \label{eq:odd-simp-rels}
    \end{align}

\section{\texorpdfstring{Exhaustive list of $p$-coefficients}{Exhaustive list of p-coefficients}}\label{app-pcoeffs}
In the main text we have explicitly shown the expressions for the $p$-coefficients in the quadratic Lagrangian~\eqref{L2_odd_Ein} for the following cases: $\textbf{Cubic}_{\text{GR-mat}}$~\eqref{eq:p-coefsGRmat},
$\textbf{Cubic}_{\text{GR-vac}}$~\eqref{eq:p-coefsGRvac}, $\textbf{Cubic}_{\text{SdS}}$~\eqref{eq:p-coefsSdS}, and $\textbf{Cubic}_{\text{Schw}}$~\eqref{eq:p-coefsSchw}.
For the cases~$\prescript{\text{SS}}{}{\textbf{Cubic}}_{\text{GR-mat}}$ and $\prescript{\text{(SS)}}{}{\textbf{Quadratic}}_{\text{GR-mat}}$, it is trivial that the $p$-coefficients are given by \eqref{eq:p-coefsGRmat}.
Here we collect the expressions for the remaining cases in Table~\ref{main-table}. Recalling that we found $p_4=0$ to be generically true in cubic HOST theories (as one would expect for theories admitting slowly-rotating black hole solutions~\cite{Mukohyama:2022skk}), here we show expressions for the remaining $p_1$, $p_2$, $p_3$, and $p_5$.

\begin{table}[H]
\centering
\stepcounter{table}
\begin{tabular}{l p{11cm} r}
\textbf{Model} & \textbf{$p$-coefficients} & \\
\hline
\begin{tabular}{lr}
$\prescript{\text{SS}}{}{\textbf{Cubic}}_{\text{GR-vac}}$, \\
$\prescript{\text{(SS)}}{}{\textbf{Quadratic}}_{\text{GR-vac}}$.
\end{tabular} &
\begin{tabular}{l}
$p_1=r^2\sqrt{1-A}(j^2-2)F_2$, \\
$p_2=-\frac{r^2}{\sqrt{1-A}}(j^2-2)F_2$, \\
$p_3=\frac{r^4}{\sqrt{1-A}}F_2$, \\
$p_5=0$. 
\end{tabular}  &
\refstepcounter{tableeqn} (\thetableeqn)\label{eq:p-coefsGroup1}\\
\hline
\begin{tabular}{l}
$\prescript{\text{SS}}{}{\textbf{Cubic}}_{\text{SdS}}$, \\
$\prescript{\text{(SS)}}{}{\textbf{Quadratic}}_{\text{SdS}}$,\\
$\prescript{\text{(SS)}}{}{\textbf{Quadratic}}_{\text{Schw}}$.
\end{tabular} &
\begin{tabular}{l}
$p_1=r^2\sqrt{1-A}(j^2-2)(F_2+2X_0A_1)$, \\
$p_2=-\frac{r^2}{\sqrt{1-A}}(j^2-2)F_2$, \\
$p_3=\frac{r^4}{\sqrt{1-A}}(F_2+2X_0A_1)$, \\
$p_5=0$. 
\end{tabular}  &
\refstepcounter{tableeqn} (\thetableeqn)\label{eq:coefsGroup2}\\
\hline
\begin{tabular}{l}
$\prescript{\text{SS}}{}{\textbf{Cubic}}_{\text{Schw}}$
\end{tabular}  &
\begin{tabular}{l}
$p_1=r\sqrt{1-A}(j^2-2)\Big[r(F_2+2X_0A_1)-81\sqrt{2X_0^3(1-A)}\,B_1\Big]$, \\
$p_2=-\frac{r^2}{\sqrt{1-A}}(j^2-2)F_2$, \\
$p_3=\frac{r^4}{\sqrt{1-A}}(F_2+2X_0A_1)$, \\
$p_5=0$. 
\end{tabular}  &
\refstepcounter{tableeqn} (\thetableeqn)\label{eq:coefsGroup3}\\
\hline
\end{tabular}
\caption{List of $p$-coefficients in the quadratic Lagrangian~\eqref{L2_odd_Ein} for odd-parity perturbations.}
\end{table}

%%%%%%%%%%%%%%%%%%%%%%%%%%%%%%%%%%%%%%%%%%%%%%%%%%
{}
\bibliographystyle{utphys}
\bibliography{StealthBH_cubic}

\end{document}